\documentclass[12pt]{article}
\usepackage[utf8]{inputenc}

\usepackage{amscd}
\usepackage{amsfonts}
\usepackage{amsmath}
\usepackage{amssymb}
\usepackage{amsthm}
\usepackage{accents}
\usepackage{youngtab}
\usepackage{braket}
\usepackage[dvipdfmx]{graphicx}
\usepackage[svgnames]{xcolor}
\usepackage{ytableau}
\usepackage{here}
\usepackage{comment}
\usepackage{tikz}
 \usetikzlibrary{patterns,arrows,calc}
\usepackage{feynmf}
\usepackage{hyperref}
\usepackage{mathtools}
\usepackage{pb-diagram}
\usepackage{simplewick}
\usepackage{comment}
\usepackage{ascmac}

\newcommand{\ba}{\begin{eqnarray}}
\newcommand{\ea}{\end{eqnarray}}
\newcommand{\bq}{\begin{equation}}
\newcommand{\eq}{\end{equation}}

\newcommand{\nn}{\nonumber}

\makeatletter
\@addtoreset{equation}{section}

\makeatother
\begin{document}

\begin{titlepage}
\vspace*{1cm}

\begin{center}
{\huge $q$-Deformation of Corner Vertex Operator Algebras by Miura Transformation}
\par
\vspace{1cm}
{\Large Koichi Harada, Yutaka Matsuo$^\dagger$, Go Noshita and Akimi Watanabe}
\\[.6cm]
{\it Department of Physics, The University of Tokyo}\\
{\it 7-3-1 Hongo, Bunkyo-ku, Tokyo 113-0033, Japan}\\
\vspace{1mm}
{\it $^\dagger$
Department of Physics \& Trans-scale Quantum Science Institute\\ \& Mathematics and Informatics Center, University of Tokyo,\\
Hongo 7-3-1, Bunkyo-ku, Tokyo 113-0033, Japan}
\\[.4cm]
\texttt{E-mail: harada, matsuo, noshita, awatanabe at hep-th.phys.s.u-tokyo.ac.jp}

\end{center}
\vspace{3cm}

\begin{abstract}
	\noindent
	Recently, Gaiotto and Rapcak proposed a generalization of $W_N$ algebra by considering the symmetry at the corner of the brane intersection (corner vertex operator algebra).  The algebra, denoted as $Y_{L,M,N}$, is characterized by three non-negative integers $L, M, N$. It has a manifest triality automorphism which interchanges $L, M, N$, and can be obtained as a reduction of $W_{1+\infty}$ algebra with a ``pit" in the plane partition representation.
	Later, Prochazka and Rapcak proposed a representation of $Y_{L,M,N}$ in terms of $L+M+N$ free bosons by a generalization of Miura transformation, where they use the fractional power differential operators.
	
	In this paper, we derive a $q$-deformation of the Miura transformation. It gives a free field representation for $q$-deformed $Y_{L,M,N}$, which is obtained as a reduction of the quantum toroidal algebra.
	We find that the $q$-deformed version has a ``simpler" structure than the original one because of the Miki duality in the quantum toroidal algebra. For instance, one can find a direct correspondence between the operators obtained by the Miura transformation and those of the quantum toroidal algebra. Furthermore, we can show that the both algebras share the same screening operators.
	\vspace{0.5cm}
\end{abstract}

\vfill

\end{titlepage}
\vfil\eject

\setcounter{footnote}{0}
\tableofcontents
\section{Introduction}
W-algebra is an extended conformal symmetry that includes the higher spin currents. It played an essential role in understanding the statistical mechanics and quantum gravity in two dimensions and rapidly developed around '90.
Recently, W-algebra begins to attract a renovated interest both from physicists and mathematicians. One of the reasons is the discovery of AGT correspondence \cite{Alday2010}, which suggests the algebra and its representation theory is critically important to understand higher dimensional gauge/string theories.  

A series of studies brought us a novel insight into W-algebras. For example, it was found that $W_{1+\infty}$ is equivalent to affine Yangian of $\mathfrak{gl}_1$ \cite{Tsymbaliuk:2014,Prochazka:2015deb}. The duality between two algebra provided us a new perspective on the study of $W_N$ algebras and was helpful to understand how 4D/2D duality works. 

Recently, Gaiotto and Rapcak \cite{Gaiotto:2017euk} found a new family of $W$-algebras from D3-branes attached to a 5-brane junction. They denoted their algebra as $Y_{L,M,N}[\Psi]$ where $L,M, N$ are the number of D3 branes and $\Psi$ is the coupling of $N=4$ super Yang-Mills in four dimensions. 
The authors of \cite{Gaiotto:2017euk, Prochazka:2017qum} also claimed that the algebra can be obtained as a truncation of the plane partition representation of the affine Yangian with a ``pit" at the location $(L, M, N)$, \cite{bershtein2018plane}. The algebras are referred to as corner VOAs or Y-algebras. 
One of the salient features of $W_{1+\infty}$ is the triality \cite{Gaberdiel:2012ku}. The corner VOA inherits it as the freedom to reshuffle three integers $L, M, N$ together with an appropriate transformation of the coupling $\Psi$.
A mathematical proof is given in \cite{creutzig2020trialities}.

As an independent study of the generalization of the $W_N$ algebra, the authors of \cite{Litvinov:2016mgi} classified the generic chiral algebras which contain spin three current. Their strategy is to use the free boson oscillators and a set of screening operators realized by the vertex operators. They derived a classification on the set of screening operators to have spin three current to commute with screening currents. 

In \cite{Prochazka:2018tlo}, Prochazka and Rapcak proposed a free field realization of the corner VOAs.  It is one of the most powerful techniques to understand W-algebras and their representation theories. A typical example is Fateev-Lukyanov's construction \cite{Fateev1988}, where  $W_N$ algebras were introduced. The explicit form of the generators is determined by the so-called Miura transformation, which is given by an $N$-th order differential operator. We note that $W_N$ algebra together with $U(1)$ current is a special example of the corner VOA with $(L,M,N)=(0,0,N)$. To obtain the generalization for the arbitrary $(L, M, N)$, Prochazka and Rapcak introduced three types of differential operators so that they will respect the triality. The new feature in this construction is the appearance of pseudo-differential operators, which involve fractional powers of derivatives. They also found that the free field realization defined in this way is consistent with the screening charges derived in \cite{Litvinov:2016mgi} by checking the commutativity at low orders. 

An interesting unexplored subject in the corner VOA is the $q$-deformation\footnote{After the corner VOA appeared, several generalizations to a broader class of W-algebras have been proposed \cite{Prochazka:2017qum, Eberhardt:2019xmf, Rapcak:2019wzw}. 
}.
The first attempt to apply $q$-deformation to the two-dimensional conformal field theory was made in \cite{Shiraishi:1995rp}, where the $q$-deformation of Virasoro algebra was proposed. It was soon generalized to $q$-deformed $W_N$ algebra, whose generators are defined by $q$-deformed Miura transformation \cite{Awata:1995zk, Feigin:1995sf}. 

In this paper, we derive a $q$-deformed version of the generalized Miura transformation in \cite{Prochazka:2018tlo}. It gives an explicit free field realization of $q$-deformed $Y_{L,M,N}$.

As we wrote, the corner VOA $Y_{L,M,N}$ is obtained as a reduction of $W_{1+\infty}$, or the affine Yangian. One can obtain the $q$-deformed $Y_{L,M,N}$ from the quantum toroidal algebra $\mathfrak{gl}_1$ (also known as Ding-Iohara-Miki algebra \cite{Ding:1996mq,Miki2007}).
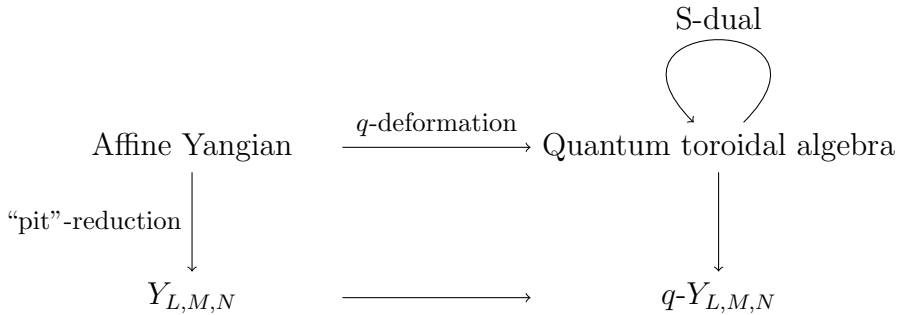
\begin{figure*}[!h]
\centering
\begin{tikzpicture}
\node(A) at (-3.5,1) {Affine Yangian};
\node(B) at (3.5,1) {Quantum toroidal algebra};
\node(C) at (-3.5,-1) {$Y_{L,M,N}$};
\node (D) at (3.5,-1) {$q$-$Y_{L,M,N}$};
\draw[->] (-1.5,1)-- node[above]{\footnotesize{$q$-deformation}} (1,1);
\draw[->] (A)--node[left]{\footnotesize{``pit"-reduction}}(C);
\draw[->](-1.5,-1)--(1,-1);
\draw[->] (B)--(D);
\draw[->](B) to [out=45,in=135,looseness=8]node[above]{S-dual} (B);
 \end{tikzpicture}
\end{figure*}

It implies several advantages of considering $q$-deformed Y-algebras by comparing its universal symmetries--the first line in the above diagram. 
First, the coproduct becomes simpler, and it allows us to deal with the tensor product representation spaces. 
Second, the quantum toroidal $\mathfrak{gl}_1$ is more symmetric. It has $SL(2,\mathbb{Z})$ symmetry (Miki-automorphism), which the affine Yangian does not share. It helps to write the free boson representation of the Drinfeld currents, and one may compare it directly with that of $q$-W algebra or $q$-$Y_{L,M,N}$ algebra. Finally, the quantum toroidal $\mathfrak{gl}_1$ has a universal R-matrix, which helps us understand that the connection with the integrable models \cite{Feigin:2015raa}. 

We utilize the benefits of the toroidal algebra in our free field construction. We first introduce three types of difference operators and define the $q$-$Y_{L,M,N}$ algebras by $q$-deformed Miura transformation. While the general form of higher spin currents in \cite{Prochazka:2018tlo} are difficult to manage, the $q$-deformation is expressed more systematically and easier to handle. In particular, it enables us to calculate the commutativity with screening charges. We prove that all the generators in our $q$-W algebras commute with the screening charges derived in \cite{bershtein2018plane}. We also discuss the relation to the quantum toroidal $\mathfrak{gl}_1$. Because the screening charges in \cite{bershtein2018plane} were originally derived in the context of the quantum toroidal $\mathfrak{gl}_1$, it should reproduce the $q$-W algebra defined by $q$-deformed Miura transformation. We found how we can reproduce all the generators from the Drinfeld currents of the quantum toroidal $\mathfrak{gl}_1$. 

We organize the paper as follows.
In section \ref{sec:2}, we review the quantum Miura transformation of $Y_{L,M,N}$ and the screening current associated with it. We also comment on the singular nature of $Y_{N,N,N}$, which should describe a null system. 
In section \ref{sec:3}, we present the $q$-deformation of the Miura transformation. We use the $q$-Pochhammer to propose the fractional $q$-difference operator (\ref{qMiuraOp}), and define the Miura transformation for $q$-deformed $Y_{L,M,N}$. As the first consistency check, we compare the generators obtained by the Miura transformation with the result of \cite{Kojima2019}, where the free boson representation of $\mathcal{W}_{q,t}(\mathfrak{sl}(2|1))$ was studied. We note that the algebra is identical to $q$-deformed $Y_{0,1,2}$ once we subtract the $U(1)$ factor. 
In section \ref{sec:4}, we establish a direct correspondence between the Miura transformation and the Drinfeld currents in the quantum toroidal $\mathfrak{gl}_1$. We note that there is no free boson representation for the Drinfeld current in the degenerate limit, and the direct comparison with the Miura transformation is not possible, which makes it difficult to show the commutativity with the screening charges. On the other hand, that becomes almost trivial once we have established the connection to the quantum toroidal $\mathfrak{gl}_1$ because its screening charge was already obtained in \cite{bershtein2018plane}. That is one of the main advantages of considering $q$-deformed version. 
In the appendix, we presented our partial proof of the $fTT$ relation proposed in section \ref{sec:3}.

\section{Miura transformation for the corner vertex operator algebra \texorpdfstring{$Y_{L,M,N}$}{YLMN}}\label{sec:2}
In this section, we review the free boson realization of the Y-algebra $Y_{L,M,N}[\Psi]$. It is a generalization of the quantum Miura transformation which defines $W_N$ algebras \cite{Fateev1988}. 

We note that the explicit form of the $W_N$-algebra is not known except for the simplest cases. Nevertheless, the free field realization of the higher spin charges and the screening currents are enough to construct the representation theory. 

The free field realization of $Y_{L,M,N}[\Psi]$ plays a similar role in the representation theory.
The $Y_{L,M,N}[\Psi]$ algebra contains $W_N$-algebra as $W_N\oplus\, U(1)\simeq Y_{0,0,N}$. Recently, the authors of \cite{Prochazka:2018tlo} proposed a generalized Miura transformation for  $Y_{L,M,N}[\Psi]$ in terms of $L+M+N$ free bosons. On the other hand, the screening currents associated with it was developed independently \cite{Litvinov:2016mgi} in the analysis of a generalization of $W_3$ algebra. Since both the definitions of Miura transformation and the screening charges are essential for us, we summarize these developments here.
\subsection{Free field realization of \texorpdfstring{$W_N$}{WN} and \texorpdfstring{$Y_{0,0,N}\simeq W_N\oplus\, U(1)$}{Y00N~WN+U(1)}}
We first review the free field realization of the $W_N$ algebra and $Y_{0,0,N}$. The algebra $W_N$ consists of spin $\ell$ chiral current $U_\ell$ with $\ell=2,3,\cdots, N$. The precise definition of the algebra $Y_{0,0,N}$ will be reviewed in the next subsection. For the moment, what we need to know is that $Y_{0,0,N}$ consists of $W_N$ algebra with extra spin 1 current.

\paragraph{Miura transformation}
We first explain the Miura transformation for $Y_{0,0,N}$ since it is simpler.
We introduce $N$ free boson fields $\phi_i$ ($i=1,\cdots, N$) with OPE,
\begin{eqnarray}\label{phiOPE}
	\phi_i(z)\phi_j(w)\sim -\delta_{ij}\log(z-w) \,.
\end{eqnarray}
The quantum Miura transformation is
\begin{equation}\label{YMiura}
	R=R_1\cdot R_2\cdots R_N=(\alpha_0\partial+\partial\phi_1(z))(\alpha_0\partial+\partial\phi_2(z))\cdots(\alpha_0\partial+\partial\phi_N(z))=\sum_{k=0}^N U_k(z)(\alpha_0\partial)^{N-k},
\end{equation}
with $R_i=\alpha_0\partial+\partial\phi_i$.
It defines the spin $\ell$ currents $U_\ell\;(\ell=0, 1, \cdots, N)$. 
The first few terms are
\begin{align}
	U_0&=1,\\
	U_1&=\sum_{i=1}^N \partial\phi_i(z),\\
	U_2&=\sum_{i<j}\partial\phi_i\partial\phi_j+\alpha_0\sum_i (i-1)\cdot\partial^2\phi_i,\\
	U_3&=\sum_{i<j<k} \partial\phi_i\partial\phi_j \partial\phi_k+\alpha_0\sum_{i<j} (i-j)\partial(\partial\phi_i\partial\phi_j)\notag\\
	&\quad+\alpha_0\sum_{i<j}(j-i-1)\partial\phi_i\partial^2\phi_j+\frac{\alpha_0^2}{2}\sum_{i}(i-1)(i-2)\partial^3\phi_i(z).
\end{align}
We identify the stress-energy tensor of $Y_{0,0,N}$ as,
	\begin{align}
	T(z)= U_2-\frac12:(U_1)^2:-\frac{N+1}{2}\alpha_0 \partial U_1 =-\frac{1}{2}:(\partial\vec\phi)^2:+\alpha_0\,\vec\rho\cdot\partial^2\vec\phi\,,\quad
	(\vec\rho)_i =i-\frac{N+1}{2}\,,
\end{align}
where $\vec\phi=(\phi_1,\cdots,\phi_N)$.
The central charge of the system is,
\begin{eqnarray}
	c=N+12\alpha_0^2\,\vec\rho\cdot\vec\rho=(N-1)(1+\alpha_0^2 N(N+1))+1\,.
\end{eqnarray}
This is the standard central charge for $W_N$ algebra plus 1.
The extra central charge comes from the nonvanishing $U(1)$ current $U_1$.

The Miura transformation for $W_N$ algebra is obtained from (\ref{YMiura}) by noting that,
\begin{align}
	R_i = &\alpha_0 \partial+\partial\phi_i = \alpha_0 \partial+\vec\epsilon_i\cdot \partial\vec\phi +\frac{1}{N}  U_1=e^{-\frac{\phi}{N\alpha_0}}\tilde R_i e^{\frac{\phi}{N\alpha_0}}\,,\\
	\tilde R_i =&\alpha_0 \partial+\vec\epsilon_i\cdot \partial\vec\phi\,.
\end{align}
Here we define $\vec\epsilon_i = \vec e_i -\frac{1}{N}\sum_i \vec e_i$ with $(\vec e_i)_j=\delta_{ij}$, and
$ \phi=\sum_{i=1}^N \phi_i$. We rewrite (\ref{YMiura}) to define the Miura transformation for $W_N$,
\begin{align}
	R&=:R_1\cdots R_N:=:e^{-\frac{\phi}{N\alpha_0}}\tilde{R}_1 e^{\frac{\phi}{N\alpha_0}}\cdots e^{-\frac{\phi}{N\alpha_0}}\tilde{R}_N e^{\frac{\phi}{N\alpha_0}}:=e^{-\frac{\phi}{N\alpha_0}}\tilde Re^{\frac{\phi}{N\alpha_0}},\label{YtoW}\\
	\tilde R &\equiv :\tilde R_1 \cdots \tilde R_N:=\sum_{k=0}^N \tilde U_k(z)(\alpha_0\partial)^{N-k}\,.
\end{align} 
The first few terms are
\begin{align}
	\tilde U_0&=1,\\
	\tilde U_1&=\sum_{i=1}^N \vec\epsilon_i\cdot\partial\vec{\phi}(z)=0,\\
	\tilde U_2&=\sum_{i<j}:(\vec\epsilon_i\cdot\partial\vec{\phi})(\vec\epsilon_j\cdot\partial\vec{\phi}):+\alpha_0\sum_i (i-1)\vec\epsilon_i\cdot\partial^2\vec{\phi}(z)=T(z)-\frac{1}{2N}:(\partial\vec{\phi})^2:=\tilde T(z)\,.
\end{align}
The $U(1)$ current is removed and
$\tilde T(z)$ generates the Virasoro algebra of the central charge 
\begin{equation}
	c=(N-1)(1+\alpha_0^2 N(N+1))\,.
\end{equation}
$\tilde R$ is precisely the Miura transformation in \cite{Fateev1988}.

\paragraph{Screening currents}
The screening current for $Y_{0,0,N}$ (and the associated $W_N$) is defined as the integral of the vertex operator which commutes with all the higher spin currents defined by the Miura transformation. It is defined using the neighboring bosonic fields as,
\begin{equation}
	\mathcal{S}_i=\oint \frac{dz}{2\pi i}:\exp(a_i\phi_i(z)-b_i\phi_{i+1}(z)):,\qquad i=1,\cdots, N-1\,.
\end{equation}
We note that the part in the Miura transformation which may have the nonvanishing OPE with $\mathcal{S}_i$ is restricted to,
\begin{align}
	R_i R_{i+1}&=\sum_{i=0}^2 U_i(z)(\alpha_0\partial)^n,\\
	&U_0=1, \quad U_1=\partial\phi_i+\partial \phi_{i+1}, \quad U_2=\partial\phi_i\partial\phi_{i+1} +\alpha_0 \partial^2\phi_{i+1}\,,
\end{align}
since other parts do not contain $\phi_i$ and $\phi_{i+1}$. In order to have vanishing commutation relation with $U_1$ and $U_2$, one determines,
\begin{equation}\label{sW}
	a_i=-b_i= b,\quad\mbox{or}\quad \frac{1}{b}\,,
\end{equation}
if we parametrize $\alpha_0=-b-1/b$. The symmetry $b\leftrightarrow 1/b$ is known as Feigin-Frenkel duality\cite{feigin1991duality}.
We note that the screening current for $W_N$ takes exactly the same form \cite{Fateev1988} since the screening currents thus determined automatically commute with the vertex operator $\exp\left(\phi/N\alpha_0\right)$ in the similarity transformation (\ref{YtoW}).

\subsection{Affine Yangian of \texorpdfstring{$\mathfrak{gl}_1$}{gl1} and \texorpdfstring{$Y_{L,M,N}$}{YLMN}}
There are two definitions of the corner vertex operator algebra. Original one is obtained by combining the Drinfeld-Sokolov reduction of super-Lie algebras and the coset construction \cite{Gaiotto:2017euk}.  The second one is the reduction of the plane partition representation of the affine Yangian of $\mathfrak{gl}_1$ \cite{Prochazka:2017qum}. For our purpose, it is more convenient to use the second.

\paragraph{Affine Yangian}
The affine Yangian of $\mathfrak{gl}_1$ is described by Drinfeld currents
\begin{equation}\label{Drinfeld}
	e(u)=\sum_{j=0}^{\infty}\frac{e_j}{u^{j+1}},\qquad
	f(u)=\sum_{j=0}^{\infty}\frac{f_j}{u^{j+1}},\qquad
	\psi(u)=1+\sigma\sum_{j=0}^{\infty}\frac{\psi_j}{u^{j+1}}.
\end{equation}
The parameter $u$ is the spectral parameter.
We follow the notation in \cite{Prochazka:2015deb} where
the algebra is  parametrized by three numbers $h_1, h_2, h_3\in \mathbb{C}$ with a constraint,
\begin{equation}
	\label{eq:sumh}
	h_1+h_2+h_3=0 .
\end{equation}
We denote $\sigma=h_1 h_2 h_3$.
The defining relations of the affine Yangian of $\mathfrak{gl}_1$ is as follows: 
\begin{equation}
	\begin{split}
		e(u)e(v)&\sim\varphi(u-v)e(v)e(u),\qquad
		f(u)f(v)\sim\varphi(v-u)f(v)f(u),\\
		\psi(u)e(v)&\sim\varphi(u-v)e(v)\psi(u),\qquad
		\psi(u)f(v)\sim\varphi(v-u)f(v)\psi(u),
	\end{split}
\end{equation}
\begin{equation}
	[\psi_i,\psi_j]=0,\quad[e_i,f_j]=\psi_{i+j},
\end{equation}
\begin{equation}
	\begin{split}
		&[\psi_0,e_j]=0,\qquad[\psi_1,e_j]=0,\qquad[\psi_2,e_j]=2e_j,\\
		&[\psi_0,f_j]=0,\qquad[\psi_1,f_j]=0,\qquad[\psi_2,f_j]=-2f_j,
	\end{split}
\end{equation}
together with Serre relations.
$\varphi(u)$ is the structure function,
\begin{equation}\label{StrFun}
	\varphi(u)=\frac{(u+h_1)(u+h_2)(u+h_3)}{(u-h_1)(u-h_2)(u-h_3)}.
\end{equation}
We note that "$\sim$" implies both sides are equal up to regular terms at $u=0$ or $v=0$.  $\psi_0$ is the central element of the algebra.
The structure function (\ref{StrFun}) is invariant under the scale transformation $h_i\rightarrow \gamma h_i$, $\psi_0\rightarrow \gamma^{-2}\psi_0$, $u\rightarrow \gamma u$. It implies that we have two independent parameters.

Proch\'azka \cite{Prochazka:2014gqa}  introduced new parameters $\lambda_i\in \mathbb{C}$ ($i=1,2,3$) which is convenient to describe the null states associated with the plane partition. They are related to $h_i$ and $\psi_0$ by, 
\begin{eqnarray}
	\lambda_i=-\frac{\psi_0 \sigma}{h_i}\,.
\end{eqnarray}
This combination is invariant under the scale transformation. The relation (\ref{eq:sumh}) is replaced by
\begin{align}\label{eq:sumhlam}
\sum_{i=1}^3 \lambda_i^{-1}=0.
\end{align}
Use of $\lambda_i$ with the constraint (\ref{eq:sumhlam}) is more symmetric representation of the two parameters in the algebra.

\paragraph{Relation with $W_{1+\infty}$}
We remind the reader of the relation between 
$W_{1+\infty}$-algebra \cite{Gaberdiel:2012ku,Gaberdiel:2011wb} and the affine Yangian. $W_{1+\infty}$-algebra consists of an infinite chiral currents $W^{(n)}(z)$ with $n=1,2,\cdots$ where $W^{(2)}$ is the stress-energy tensor, $W^{(1)}(z)$ is a free $U(1)$ current and $W^{(n)}(z)$ ($n>2$) are the higher spin currents. One may define it as a $W_N$-algebra with $N\rightarrow \infty$ limit with an extra decoupled $U(1)$ current and an extra deformation parameter. Each current has an expansion $W^{(s)}(z)=\sum_{n\in \mathbb{Z}} W^{(s)}_n z^{-s-n}$.
See also \cite{linshaw2017universal} for $W_\infty$.

$W_{1+\infty}$ looks very different from the affine Yangian while they are equivalent. Roughly speaking, the correspondence between the generators is
\begin{equation}
	W^{(s)}_{-1}\leftrightarrow e_{s-1},\quad
	W^{(s)}_{0}\leftrightarrow \psi_{s-2},\quad
	W^{(s)}_{1}\leftrightarrow f_{s-1}.
\end{equation}
In general, the correspondence is much more involved, see, for instance, \cite{schiffmann2013cherednik}.\footnote{The basic idea is as follows. The identification of two algebra can be made through three generators $e_0\sim W^{(1)}_{-1}$, $f_0\sim W^{(1)}_1$, $\psi_1\sim W^{(3)}_0$. One may generate the whole algebra by taking commutator of these elementary generators.} In a sense, the expansion (\ref{Drinfeld}) is with respect to the spin (upper) index of $W^{(s)}_n$, while the expansion in $W_{1+\infty}$ is with respect to the lower index. There is an obvious asymmetry between them, the upper index goes from $1$ to $\infty$, while the lower index can take arbitrary integer.

The algebra contains two parameters, $c$ and $x$, where $c$ is the central charge of Virasoro algebra and $x$ is the parameter which describes the OPE coefficients of higher currents. The relations between the parameters of two algebras are \cite{Prochazka:2014gqa},
\begin{equation}
\label{eq:Winftyparameter}
c=1+\prod_{i=1}^3 (\lambda_i-1), \quad
x^2=144(c+1) \prod_{i=1}^3 (\lambda_i-2)(\lambda_i-3)^{-1}\,.
\end{equation}
As we see, the reduction to $Y_{L,M,N}$ is easiest to describe in the affine Yangian. On the other hand, the free boson representation of $Y_{L,M,N}$ describes higher spin currents of $W_{1+\infty}$. This is the origin of the complication to define the Miura transformation.

\paragraph{Representation by plane partition and the reduction to $Y_{L,M,N}$:}

In the following, we consider a representation by the affine Yangian in terms of a plane partition \cite{feigin2012quantum, Prochazka:2015deb}. We introduce a set of basis with a label of a plane partition $\Lambda$, and it spans the Hilbert space of the algebra.
The operator $\psi_i$ is diagonal with respect to $|\lambda\rangle$ and $e_i$($f_i$) play a role of adding (removing) a box to $\Lambda$:
\begin{eqnarray}
	\psi(u)\ket{\Lambda}&=&\psi_{\Lambda}(u)\ket{\Lambda},\\
	\label{eq:addbox}
	e(u)\ket{\Lambda}&=&\sum_{\raisebox{1.7pt}{\fbox{}}\in\Lambda^+}\frac{1}{u-h_{{\fbox{}}}}\sqrt{-\frac{1}{\sigma}{\rm res}_{u\to q+h_{\fbox{}}}\psi_{\Lambda}(u)}\ket{\Lambda+\raisebox{3.5pt}{\fbox{}}},\\
	\label{eq:removebox}
	f(u)\ket{\Lambda}&=&\sum_{\raisebox{1.7pt}{\fbox{}}\in\Lambda^-}\frac{1}{u-h_{{\fbox{}}}}\sqrt{-\frac{1}{\sigma}{\rm res}_{u\to q+h_{\fbox{}}}\psi_{\Lambda-\raisebox{2.2pt}{\fbox{}}}(u)}\ket{\Lambda-\raisebox{3.5pt}{\fbox{}}},
\end{eqnarray}
and
\begin{equation}
	\label{eq:psieigenvalue}
	\begin{split}
		&\psi_{\Lambda}(u)=\psi_0(u)\prod_{\raisebox{1.5pt}{\fbox{}}\in\Lambda}\varphi(u-h_{\fbox{}}),\\
		&\psi_0(u)=\frac{u+\psi_0\sigma}{u}.
	\end{split}
\end{equation}
Here, $\Lambda^{\pm}$ are the places where we can add (or remove) boxes so that the shape of plane partition is consistent.
We introduce a coordinate for each box in the plane partition. We assign $(0,0,0)$ to the origin of the partition and $(x_1,x_2,x_3)\in (\mathbb{Z}_{\geq 0})^{\otimes 3}$ for a general box. We assign
\begin{equation}
	\label{eq:assignedvalue}
	h_{\fbox{}}=h_1x_1+h_2x_2+h_3x_3
\end{equation}
to the box located at $(x_1,x_2,x_3)$. 

To define $Y_{L,M,N}$, we impose $\lambda_i$ to satisfy an extra condition:
\ba\label{DegLMN}
\frac{L}{\lambda_1}+\frac{M}{\lambda_2}+\frac{N}{\lambda_3}=1.
\ea
From the explicit representation (\ref{eq:addbox},\ref{eq:removebox}), one can derive the basis $|\Lambda\rangle$ which contains a box with a coordinate $(L, M, N)$ becomes null. In \cite{bershtein2018plane}, the authors analyzed the similar problem for the quantum toroidal algebra. They describe the restriction to the plane partition realization as ``pit" at $(L, M, N)$.
In \cite{Prochazka:2017qum}, the authors claimed that the affine Yangian whose parameter is constrained by this condition is equivalent to the vertex operator algebra $Y_{L,M,N}[\Psi]$ in \cite{Gaiotto:2017euk}. The parameter $\Psi$ of the algebra is written as,
\begin{equation}\label{def:Psi}
	\Psi=-\lambda_1/\lambda_2\,.
\end{equation}
One may solve (\ref{eq:sumhlam}, \ref{DegLMN},\ref{def:Psi}) to express $\lambda_i$ as,
\begin{align}
	\lambda_1&=L-M\Psi+N(\Psi-1),\\
	\lambda_2&=-\frac{L}{\Psi}+M+N\left(\frac{1}{\Psi}-1\right),\\
	\lambda_3&=\frac{L}{\Psi-1}+\frac{M\Psi}{1-\Psi}+N\,.
\end{align}

We note that the condition (\ref{DegLMN}) has a shift symmetry, 
\begin{eqnarray}\label{shift}
	L\rightarrow L+k,\quad
	M\rightarrow M+k,\quad
	N\rightarrow N+k.
\end{eqnarray}
for $k\in \mathbb{Z}$ due to (\ref{eq:sumhlam}). It allows the redefinition the location of the pit such that the smallest elements are zero and others are greater than zero.
The character of the plane partitions with a pit was derived in \cite{bershtein2018plane}.

For the special case $Y_{0,0,N}$, we have a pit at the box $(0,0,N)$. With such condition, the height (the number of layers in $z$ direction) of the plane partition $\Lambda$ for the nonvanishing states is restricted to $N$. 
One may decompose the plane partition layer by layer into $N$-tuple Young diagrams $Y_1,\cdots, Y_N$ with the condition $Y_1\succeq\cdots \succeq Y_N$. We note that the Fock space of the free boson is written in terms of the basis labeled by Young diagrams. They give a representation space of $W_N$ algebra with an extra $U(1)$ factor, which we discussed in the previous subsection. 

\subsection{Miura operators}
In \cite{Prochazka:2018tlo}, the authors proposed the Miura transformation which generate the higher spin charges of the algebra $Y_{L,M,N}$.
It is written in terms of $L+M+N$ free boson fields $\phi_i(z)$ ($i=1,\cdots, L+M+N$) which obeys the standard OPE relation (\ref{phiOPE}).
$Y_{L,M,N}$ always contains the Virasoro current in the following standard form:
\begin{equation}\label{SETensor}
	T(z)=-\frac{1}{2}:\partial\vec\phi\cdot\partial\vec\phi:+\vec\tau \partial^2\vec\phi\,.
\end{equation}
with the central charge $c=L+M+N+12\vec\tau^2$.
The vertex operator and the conformal dimension is given by,
\begin{eqnarray}
	V_{\vec a}(z)=:\exp(\vec a\cdot\vec\phi(z)):,\qquad \Delta(\vec a)= \frac12 \vec a\cdot (2\vec\tau -\vec a)\,.
\end{eqnarray}

The Miura transformation\cite{Prochazka:2018tlo} consists of three types (pseudo-)differential operators,
which we will refer as the Miura operator
\begin{equation}\label{eq:degenerate_Miuraop}
	R^{(c)}(\phi)=:(\partial_z+\mu_c \partial_z\phi )^{\nu_c}:=:e^{-\mu_c\phi}\partial_z^{\nu_c}e^{\mu_c\phi}:, \quad c=1,2,3\,.
\end{equation} 
The parameters are set to,
\begin{align}
	\begin{array}{ll}
		\mu_1=\alpha, \quad & \nu_1=\beta^2,\\
		\mu_2=\beta, \quad & \nu_2=\alpha^{2},\\
		\mu_3=\alpha\beta, \quad & \nu_3=1,
	\end{array}
\end{align}
with $\alpha,\beta\in \mathbb{C}$ with a constraint:
\begin{equation}
	\alpha^2+\beta^2+1=0\,.
\end{equation}
They are related to the $\Psi$ parameter in $Y_{L,M,N}[\Psi]$ as,
\begin{equation}
	\Psi=-\alpha^2/\beta^2\,,\quad\mbox{or}\quad \alpha^2=\frac{\Psi}{1-\Psi},\quad \beta^2=\frac{1}{\Psi-1}\,.
\end{equation}
These parameters are introduced here to simplify the free boson representations.
The other parameters of the affine Yangian can be set to,
\begin{align}
	& h_1=\beta/\alpha, \quad h_2=\alpha/\beta,\quad h_3=\alpha_0=\frac{1}{\alpha\beta}\,,\quad\psi_0=L\nu_1+M\nu_2+N\nu_3\,,\label{htoalpha}\\
	& \lambda_c=-\psi_0/\nu_c \quad (c=1,2,3)\label{lamtonu}\,.
\end{align}
Our definition of the Miura operator is identical to the \cite{Prochazka:2018tlo}, which we refer to as $R^{(PR:c)}$
	\begin{align}\label{eq:MiuraOp}
		&R^{(PR:c)}= :(\alpha_0\partial_z +\frac{h_3}{h_c}J^{(c)})^{h_c/h_3}:\,,\\
		&J^{(c)}(z) J^{(c)}(w) =-\frac{h_c}{\sigma} \frac{1}{(z-w)^2}\,.\label{eq:Ji}
	\end{align}
up to the overall coefficient and the rescaling of the currents by $J^{(c)}(z)=\sqrt{\frac{h_c}{\sigma}}\partial\phi$, with:
\begin{align}
		& \mu_c=\frac{1}{h_c}\sqrt{\frac{h_c}{\sigma}},\quad \nu_c=\frac{h_c}{h_3}.\label{eq:mu_nu_by_h}
\end{align}
The parameters $\mu_c, \nu_c$ satisfy the relations,
\begin{equation}
\sum_{c=1,2,3} \nu_c=0,\quad \mu_c^2=\omega/\nu_c,\quad \omega=\alpha^2\beta^{2}=(\sigma)^{-2}.
\end{equation}

\paragraph{Expansion of pseudo differential operators}
The action of the pseudo-differential operator on a given function is given by,
\begin{align}
	\partial_z^\nu f(z)= \sum_{n=0}^\infty \frac{(-1)^n (-\nu)_n}{n!}(\partial_z^n f(z)) \partial_z^{\nu-n}\,,
\end{align}
where 
\begin{equation}\label{Pochhammer}
	(\nu)_n=(\nu+n-1)\cdots (\nu+1)\nu\,,
\end{equation} 
is the Pochhammer symbol.
For the expansion of the pseudo-differential operator, the following formula is useful.
\begin{eqnarray}
	:(\partial+\mu\partial\phi)^\nu:&=&:e^{-\mu\phi}\partial^\nu e^{\mu\phi}:\nn\\
	&=&\sum_{n=0}^\infty(-1)^n \frac{(-\nu)_n}{n!} P_n[\mu\partial\phi]\partial^{\nu-n}\,,\label{MiuraExpansion}\\
	P_n[J]&=& :(\partial+J)^n:\cdot 1\,.
\end{eqnarray}
The first few examples of $P_n$ are,
\begin{equation}
	P_1[J]=J,\quad
	P_2[J]=\partial J+:J^2:,\quad
	P_3[J]= \partial^2J+:3J\partial J:+:J^3:\,.
\end{equation}
We will use a notation,
\begin{align}
	&:(\partial+\mu_c\partial\phi)^{\nu_c}: =\sum_{n=1}^\infty U^{(c)}_n \partial^{\nu_c-n}\,,\\
	&U^{(c)}_n = (-1)^n\frac{(-\nu_c)_n}{n!} P_n[\mu_c \partial \phi]\,,
\end{align}
where $U^{(c)}_n$ is the ``spin $n$" component of the expansion.

\subsection{Miura transformation}\label{s:Miura}
One defines a generalized Miura transformation for $Y_{L,M,N}$ as,
\begin{align}
	R&=R^{(c_1)}(\phi_1)R^{(c_2)}(\phi_2)\cdots R^{(c_n)}(\phi_n)=\sum_{s=0}^\infty U_s(\vec\phi)\,\partial^{\pi_1-s},\label{Miura}
\end{align}
where $n=L+M+N$. We use a notation $\pi_1=\sum_{i=1}^n \nu_{c_i}=L\nu_1+M\nu_2+N\nu_3=-\psi_0$.
We will generalize it to $\pi_s=\sum_{i=1}^n (\nu_{c_i})^s=L\nu_1^s+M\nu_2^s+N\nu_3^s$ in the following.
Among the set 
\begin{equation}
\vec{c}=\{c_1,\cdots, c_n\},\label{eq:vecc}
\end{equation}
$L$ indices equal $1$, $M$ indices equals $2$, and $N$ indices equals $3$.
The expansion coefficient $U_s(\vec\phi)$ plays the role of spin $s$ current in $Y_{L,M,N}$.
When all $\{c_1,\cdots, c_n\}$ equal $3$, we recover the original Miura transformation of $W_n$ algebra (\ref{YMiura}) up to overall constant.

The authors of \cite{Prochazka:2018tlo} proposed,
\begin{itemize}
	\item The algebra $Y_{L,M,N}$ does not depend on the order of the product of the Miura operators.
	\item There is a shift symmetry $Y_{L,M,N}\rightarrow Y_{L+1,M+1,N+1}$.
\end{itemize}
There is a partial arguments for the first point \cite{Prochazka2019}. It is based on the existence of $\mathcal{R}$-matrix, which changes the order of the Miura operators,
\begin{equation}
	\mathcal{R}_{c_1c_2}R^{(c_1)}(\phi_1)R^{(c_2)}(\phi_2)=R^{(c_2)}(\phi_2)R^{(c_1)}(\phi_1)\mathcal{R}_{c_1c_2}\,.
\end{equation}
Their argument is a generalization of \cite{Maulik:2012wi, Zhu2015} for the $R^{(3)}$ operator in the homogeneous form (\ref{YMiura}).  There is also a different reasoning \cite{Litvinov:2016mgi} based on the correlation function of the screening currents.

For the second point, the existence of the shift symmetry implies that the currents $U_r$ defined by $R^{(1)}(\phi_1)R^{(2)}(\phi_2)R^{(3)}(\phi_3)=\sum_{r=0}^\infty U_r(\vec\phi)\partial^{-r}$, which describes $Y_{1,1,1}\simeq Y_{0,0,0}$ should be null operators. We will give some arguments that this statement holds.

We come back to the expansion of (\ref{Miura}) more explicitly.
\begin{align}
	U_0=&1\,,\\
	U_1=&\sum_{i=1}^n U^{(c_i)}_1=\sum_{i=1}^n \mu_{c_i} \nu_{c_i} \partial\phi_{i} =\vec g\cdot \partial\vec\phi,\quad \vec g=(\mu_{c_1}\nu_{c_1},\cdots,\mu_{c_n}\nu_{c_n})\,,\\
	U_2=&\sum_{i=1}^n U^{(c_i)}_2+\sum_{i<j} (U^{(c_i)}_1 U^{(c_j)}_1+\nu_{c_i} \partial U^{(c_j)}_1)\,,\\
	U_3=&\sum_{i=1}^n U^{(c_i)}_3+\sum_{i<j} (U^{(c_i)}_1 U^{(c_j)}_2+U^{(c_i)}_2 U^{(c_j)}_1)+\sum_{i<j<k} U^{(c_i)}_1U^{(c_j)}_1U^{(c_k)}_1\nn\\
	&+\sum_{i<j}(\nu_{c_i} \partial U^{(c_j)}_2+(\nu_{c_i}-1) U^{(c_i)}_1\partial U^{(c_j)}_1)\nn\\ 
	& +\sum_{i<j<k}\left(\nu_{c_i}\partial(U^{(c_j)}_1U^{(c_k)}_1) +U^{(c_i)}_1\nu_{c_j}\partial U^{(c_k)}_1\right)\nn\\
	&+\sum_{i<j}\frac{\nu_{c_i}(\nu_{c_i}-1)}{2}\partial^2 U^{(c_{j})}_1+\sum_{i<j<k}\nu_{c_i}\nu_{c_j}\partial^2 U^{(c_k)}_1\,.\label{u3}
\end{align}
The generic generator $U_s$ is formally written as,
\begin{align}\label{Us}
	U_s=\sum_{\ell_1+\cdots+\ell_n=s, \,\ell_i\geq 0} \frac{\prod_{i=1}^n \left((-1)^{\ell_i}(-\nu_i)_{\ell_i}\right)}{\ell_1!\cdots \ell_n!}:e^{-\mu_{c_1}\phi_1}\partial^{\ell_1}\left(e^{\mu_{c_1}\phi_1-\mu_{c_2}\phi_2}
	\partial^{\ell_2}\left(e^{\mu_{c_2}\phi_2-\mu_{c_3}\phi_3}\partial^{\ell_3}\left(\cdots \right)\right)\right):\,.
\end{align}

Let us summarize the basic properties of CFTs defined by these operators.
\begin{itemize}
	\item Spin one $U(1)$ current: $J(z)=U_1(z)$. Their OPE is
\begin{equation}\label{JJ}
	J(z)J(w)= \frac{\omega\pi_1}{(z-w)^2}. 
\end{equation}
	\item Stress energy tensor $T(z)$: it takes of the form (\ref{SETensor}) by combining
	\begin{align}
		T_0(z)= \frac{1}{\omega}(U_2-\frac12:(U_1)^2:)=-\frac{1}{2}:(\partial\vec\phi)^2:+\vec\tau_0\cdot\partial^2\vec\phi\,,\qquad
	(\vec\tau_0)_i =\frac{\nu_{c_i}-1+2\sum_{j<i} \nu_{c_j}}{2\mu_{c_i}}\,.
	\end{align}
	We have to modify $T_0$ by adding a term proportional to $\partial J(z)$ to impose $\vec \tau\cdot \vec g=0$.\footnote{When $L=M=N$, $\vec g^2=0$ and the expression for $T(z)$ appears to be ill-defined. Actually the operator $\partial J$ becomes null operator and its coefficient is arbitrary. Such a modification does not change the central charge and the structure of the screening operators.}
	\begin{align}
		T(z)=&T_0(z)-\frac{\vec g\cdot\vec\tau_0}{\vec g^2}\partial J(z)=-\frac{1}{2}:(\partial\vec\phi)^2:+\vec\tau\cdot\partial^2\vec\phi\,,\label{TzM}\\
		(\vec\tau)_i =& \frac{\sum_{j<i}\nu_{c_j}-\sum_{j>i}\nu_{c_j}}{2\mu_{c_i}}\,.
	\end{align}
We note that
\begin{align}
	\vec g^2 &= \sum_i \mu_{c_i}^2\nu_{c_i}^2=\omega\sum_i\nu_{c_i} =\omega\pi_1\,,\\
	\vec g\cdot \tau_0&=\sum_i \beta_{c_i}\nu_{c_i} \frac{\nu_{c_i}-1+2\sum_{j<i} \nu_{c_j}}{2\beta_{c_i}}
	=\cdots=\frac12 \pi_1(\pi_1-1)\,,\\
	\vec \tau_0^2 &= \frac14\sum_i\frac{(\nu_{c_i}-1+2\sum_{j<i} \nu_{c_j})^2}{(\mu_{c_i})^2}=
	\frac{1}{4\omega}\sum_i \nu_{c_i}(\nu_{c_i}-1+2\sum_{j<i} \nu_{c_j})^2\nn\\
	& =\cdots = \frac{1}{4\omega}(-\frac13\pi_3+\frac{4}{3}(\pi_1)^3-2(\pi_1)^2+\pi_1)\,.
\end{align}
It implies that the central charge of (\ref{TzM}) is,
\begin{equation}
	c=n+12\left(
	\vec\tau_0^2-\frac{(\vec g\cdot\vec\tau_0)^2}{\vec g^2}
	\right)=n+\frac{1}{\omega}((\pi_1)^3-\pi_3)\,.\label{c}
\end{equation}
It is identical to (\ref{eq:Winftyparameter}) after some computation by using the indentities (\ref{htoalpha},\ref{lamtonu}).
\end{itemize}

\paragraph{Comments:}
\begin{enumerate}
	\item When the Miura transformation is homogeneous, namely $R^{(c_1)}(\phi)=\cdots=R^{(c_n)}(\phi)=R^{(c)}(\phi)$, the central charge becomes
	\begin{equation}
		c_c=n+\frac{\nu_c^3}{\omega}(n^3-n)\,.
	\end{equation}
	We note that
	$$
	\frac{\nu_1^3}{\omega}=(\alpha+\alpha^{-1})^2,\quad
	\frac{\nu_2^3}{\omega}=(\beta+\beta^{-1})^2,\quad
	\frac{\nu_3^3}{\omega}=(\alpha/\beta+\beta/\alpha)^2\,.
	$$
	It takes the form of the central charge of $W_n$ algebra and a $U(1)$ boson.
	As we will see, the screening currents for $Y_{N,0,0}$, $Y_{0,N,0}$, $Y_{0,0,N}$ takes the similar form.
	It implies that these three Miura transformations defines the same algebra with the parameter redefined. 
	\item It is clear from the final expression (\ref{JJ},\ref{c}) that the level of $U(1)$ (the numerator factor of (\ref{JJ})) and the central charge is independent of the order of Miura transformation (\ref{Miura}).

	\item For $Y_{1,1,1}$, which be equivalent to  $Y_{0,0,0}$, $\pi_1=0$, $\pi_3=3\omega$. It gives the level of $U(1)$ current algebra and the central charge of $W_{\infty}$ to vanish. It implies that there are no operators proportional to the identity operator in the commutation relations. From this fact, one may claim that the norm of the states generated by $U^{(s)}_{-n}$, $s,n=1,2,\cdots,\infty$ (obtained from the mode expansion of $U_s(z)$),  vanishes, while it is generated by non-trivial free bosons. Thus, the Miura transformation for $Y_{1,1,1}$ gives ``null operators" $U_s$ while their appearance is very complicated. 
\end{enumerate}

\subsection{Screening currents}
The screening currents for the Miura transformation (\ref{Miura}) should commute with the neighboring Miura operators,
$ R^{(c_i)}(\phi_i)R^{(c_{i+1})}(\phi_{i+1})$. One expands,
\begin{equation}\label{sMiura}
 R^{(a)}(\phi_1)R^{(b)}(\phi_2)=\partial^{\nu_a+\nu_b}+U_1\partial^{\nu_a+\nu_b-1}+U_2\partial^{\nu_a+\nu_b-2}+\cdots\,.
\end{equation}
We define the screening currents as
\begin{equation}
	\mathcal{S}_{ab}=\oint \frac{dz}{2\pi \mathrm{i}} :\exp(\vec k_{ab}\cdot \vec \phi(z)):, \quad
	\vec k_{ab}=(k_{ab}^1,k_{ab}^2), \quad \vec\phi=(\phi_1,\phi_2)\,.
\end{equation}
They should commute with $U_n$ in (\ref{sMiura}). We use the notation ($\vec g, \vec \tau$) in section \ref{s:Miura} for this short Miura transformation. The commutativity with $U_1$ and $U_2$ gives,
\begin{equation}\label{cond:scr}
	\vec g\cdot \vec k_{ab}=0,\quad \Delta(\vec k_{ab})=\frac12\vec k_{12}\cdot(2\vec \tau-\vec k_{ab})=1\,.
\end{equation}
The solutions to these equations are given as follows,
\begin{align}
	\vec k_{cc}^{\pm} &= \frac{\nu_{c\pm 1}}{\mu_c}(1,-1)\,,\\
	\vec k_{ab} & =\left(\frac{\nu_b}{\mu_a}, -\frac{\nu_a}{\mu_b}\right)\,,\qquad a\neq b\,.
\end{align}
More explicitly,
\begin{align}
	\vec k_{11}&= (\alpha^{\pm 1},-\alpha^{\pm 1}),\quad
	\vec k_{22}= (\beta^{\pm 1},-\beta^{\pm 1}),\quad
	\vec k_{33}= ((\alpha/\beta)^{\pm 1},-(\alpha/\beta)^{\pm 1})\,,\\
	\vec k_{12}&=(\alpha, -\beta),\quad \vec k_{21}=	(\beta, -\alpha),\quad
	\vec k_{13}=(\alpha^{-1}, -\beta/\alpha),\quad \vec k_{31}=	(\beta/\alpha, -\alpha^{-1}),\nn\\
	\vec k_{23}&=(\beta^{-1}, -\alpha/\beta),\quad \vec k_{32}=	(\alpha/\beta, -\beta^{-1})\,.
\end{align}

We note that for $R^{(c)}R^{(c)}$, we have two solutions which depends the index $a$ which is different from $c$ in the set $\{1,2,3\}$. We denote them as $c\pm 1$.
This is the reason that $\vec k_{cc}$ has an extra upper index. For the inhomogeneous case $R^{(a)}R^{(b)}$ ($a\neq b$), we have a second solution $-2\vec k_{ab}$ for (\ref{cond:scr}). The vertex operator for the second solution, however, does not commute with the higher currents.

The screening current for $\mathcal{S}_{ab}$ ($a\neq b$) anticommutes with itself. Thus it is nilpotent $(\mathcal{S}_{ab})^2=0$ \cite{bershtein2018plane}. We refer such screening charge as ``fermionic", while we refer the screening  charge $\mathcal{S}^{\pm}_{cc}$ as ``bosonic".

To examine the commutativity of the screening currents with the higher charges $U_n$ in general, we evaluate,
\begin{align}
	&\sum_{n=0}^\infty \left[\mathcal{S}_{ab}, U_n(w)\right]\partial_w^{\nu_a+\nu_b-n}\nn\\
	&=\oint_{C_w} \frac{dz}{2\pi \mathrm{i}} e^{\vec k_{ab}\cdot \vec \phi(z)}
	U_n(w)\partial_w^{\nu_a+\nu_b-n}\nn\\
	&= \mbox{Res}_{z=w}:e^{\vec k_{ab}\cdot \vec \phi(z)}:
	:R^{(a)}(w)R^{(b)}(w):\nn\\
	&=\mbox{Res}_{z=w}\left(:e^{\vec k_{ab}\cdot \vec \phi(z)}::e^{-\mu_a \phi_1(w)}\partial^{\nu_a}e^{\mu_a\phi_1(w)}\cdot e^{-\mu_b \phi_2(w)}\partial^{\nu_b}e^{\mu_a\phi_2(w)}:\right)\,.\label{cSM}
\end{align}
In order to show the screening charge commutes with arbitrary $U_n$, one has to show the residue in the last line is zero.

To go further, we note that there are some components that do not have OPE with $e^{\vec k_{ab}\phi}$.

\begin{itemize}
	\item Bosonic screening charge $a=b$: Since $\vec k_{aa}\propto (1,-1)$, we have the following decomposition.
	\begin{equation}
		\phi_\pm=\phi_1\pm \phi_2, \quad \phi_\pm(z)\phi_\pm(z)\sim -2\log(z-w),\quad
		\phi_+(z)\phi_-(w)\sim 0\,.
	\end{equation}
	We rewrite,
	\begin{align}
		:R^{(a)}R^{(a)}:&=:e^{-\frac{\mu_a}2 \phi_+} R^{(aa)}(\phi_-)e^{+\frac{\mu_a}2 \phi_+}\,,\\
		R^{(aa)}(\phi_-) &= :e^{-\frac{\mu_a}2 \phi_-}\partial^{\nu_a}e^{{\mu_a} \phi_-}\partial^{\nu_a}e^{-\frac{\mu_a}2 \phi_-}:\,.
	\end{align}
	Since the vertex operator in the screening operator commutes with $\phi_+$, the last line in (\ref{cSM}) is simplified to,
	\begin{align}\label{resb}
		:e^{-\frac{\mu_a}2 \phi_+}: \mbox{Res}_{z=w}\left(:e^{\frac{\nu_b}{\mu_a} \phi_-(z)}::R^{(aa)}(\phi_-(w)):\right):e^{\frac{\mu_a}2 \phi_+}:\,,
	\end{align}
	where we take $\vec k^{(b)}_{aa}$ ($b\neq a$) for the choice of the two screening currents.
	\item Fermionic screening charge $a\neq b$: Let $c$ be the third number other than $a,b$ in the set $(1,2,3)$. First we note some identities,
	\begin{equation}\label{identities}
		\mu_a^2+\mu_b^2=-\nu_c^2,\quad \mu_a\mu_b=\mu_c\nu_c\,.
	\end{equation}
	We redefine the free boson generator,
	\begin{align}
		&	\phi_+=\frac{\mu_b\phi_1+\mu_a\phi_2}{\nu_c},\quad
		\phi_-=\frac{\mu_a\phi_1-\mu_b\phi_2}{\nu_c}\,,\\
		& 	\phi_\pm(z)\phi_\pm(z)\sim \log(z-w),\quad
		\phi_+(z)\phi_-(w)\sim 0\,.
	\end{align}
	We derive after the help of (\ref{identities}),
	\begin{align}
		\vec\kappa_{ab}\cdot \vec\phi&=\phi_- \,,\\
		:R^{(a)}R^{(b)}:&=:e^{-\mu_c \phi_+} R^{(ab)}(\phi_-)e^{\mu_c \phi_+}:\,,\\
		R^{(ab)}(\phi_-) &= :e^{\nu_b \phi_-}\partial^{\nu_a}e^{{\nu_c\phi_-}}\partial^{\nu_b}e^{\nu_a\phi_-}:\,.
	\end{align}
	Since the vertex operator in the screening operator commutes with $\phi_+$, the last line in (\ref{cSM}) is simplified to,
	\begin{align}
		:e^{-\mu_c \phi_+(w)}: \mbox{Res}_{z=w}\left(:e^{\phi_-(z)}::R^{(ab)}(\phi_-):\right):e^{\mu_c\phi_+}(w):\,.\label{resf}
	\end{align}
	
\end{itemize}
We do not have an analytic proof the residue appearing in (\ref{resb},\ref{resf}) vanishes. It is, however, possible to calculate them by computer.  
We check the residue in (\ref{resb},\ref{resf}) vanishes up to spin 8.
In section \ref{sec:4}, the commutativity of the screening currents is explicitly proved for the $q$-deformed case. By taking the degenerate limit $q\to 1$, we obtain the analytic proof of the statement.

\section{\texorpdfstring{$q$}{q}-deformation of corner VOA} \label{sec:3}
In this section, we propose a $q$-deformation of the Miura transformation for $Y_{L, M, N }$. 

\subsection{\texorpdfstring{$q$}{q}-deformation of the Miura transformation for \texorpdfstring{$Y_{0,0,N}$}{Y0,0,N} and  \texorpdfstring{$W_N$}{WN} algebra}
As a warm-up, we first review the $q$-deformation of the Miura transformation of $Y_{0, 0, N}$ and $W_N$-algebra \cite{Shiraishi:1995rp,Awata:1995zk,Feigin:1995sf}. It is written to get accustomed to the $q$-deformation of the oscillator algebra. The readers who are familiar with the subject can skip this subsection.

\subsubsection{\texorpdfstring{$q$}{q}-deformation of the Heisenberg algebra and the Miura transformation for \texorpdfstring{$Y_{0,0,N}$}{Y00N}}
As in the previous section, the construction of the $q$-deformation of $Y_{0,0, N}$ is similar to $W_N$ algebra up to the subtraction of $U(1)$ part.
We introduce the $q$-deformed Miura operators. By expanding the product of these operators we obtain the explicit generators which generate $q$-$Y_{0,0,N}$.

$Y_{0,0,N}$ before the $q$-deformation depends on three parameters $h_1, h_2, h_3$ and they satisfy $h_1+h_2+h_3=0$.
Let the parameters of the $q$-deformation be $q_1,q_2,q_3$.
$h_c\;(c=1,2,3)$ and $q_c$ are related with\footnote{The parameter $\epsilon$ can be absorbed by rescaling $h_i$. However, we keep $\epsilon$ in order to take the degenerate limit.}
\begin{align}
    q_c=e^{\epsilon h_c},\label{eq:rel_qc_hc}
\end{align}
where $\epsilon$ is a constant.
They have the following relation corresponding to $h_1+h_2+h_3=0$,
\begin{align}
    q_1q_2q_3=1.
\end{align}

We introduce a $q$-deformation of the $N$ free bosons defined by the Heisenberg algebra \cite{bershtein2018plane},
\begin{align}
    [a_m^{(i)},a_n^{(j)}]=-m\frac{(q_3^\frac{m}{2}-q_3^{-\frac{m}{2}})^3}{\kappa_m}\delta_{m+n,0}\delta_{i,j},\quad (i=1,2,\cdots, N)\label{eq:Y00N_q_boson}
\end{align}
where $\kappa_m$ is defined as
\begin{align}
\kappa_m= \prod_{c=1}^3 (q_c^{\frac{m}{2}}-q_c^{-\frac{m}{2}}).\label{eq:kappa}
\end{align}

In the $q\rightarrow 1$ (or $\epsilon\to 0$) limit, which we will refer to as the degenerate limit,
the coefficient in the right-hand side of (\ref{eq:Y00N_q_boson}) becomes
\begin{align}
    -m\frac{(q_3^\frac{m}{2}-q_3^{-\frac{m}{2}})^3}{\kappa_m}
    \to
    -m\frac{h_3^2}{h_1h_2}+\mathcal{O}(\epsilon).
\end{align}
This is the same as (\ref{eq:Ji}).

We define vertex operators $\Lambda_i$ by using free bosons $a_r^{(i)}$ as
\begin{align}
    \Lambda_i(z)
    &=\exp\left(\epsilon\frac{h_1h_2}{h_3} a_0^{(i)}\right)
    \exp\left[
    \sum_{r>0}\frac{\kappa_r}{r} q_3^{-\frac{r}{2}N}
    \frac{q_3^{\frac{r}{2}(i-1)}}{(q_3^\frac{r}{2}-q_3^{-\frac{r}{2}})^2}
    a_{-r}^{(i)} z^r\right]\notag\\
    &\quad\times \exp\left[
    \sum_{r>0}\frac{\kappa_r}{r} q_3^{\frac{r}{2}N} \frac{q_3^{-\frac{r}{2}i}}{(q_3^\frac{r}{2}-q_3^{-\frac{r}{2}})^2}
    a_r^{(i)} z^{-r}
    +\sum_{r>0}-\frac{\kappa_r}{r}q_3^{\frac{r}{2}N}
    \sum_{j=i+1}^N \frac{q_3^{-\frac{r}{2}(j-1)}}{q_3^\frac{r}{2}-q_3^{-\frac{r}{2}}}
    a_r^{(j)} z^{-r}
    \right],\label{eq:Y00N_Lambda_i}
\end{align}
The second term in the second line describes the mixing of the off-diagonal ($j\neq i$) components of the free bosons. It comes from the coproduct structure of the quantum toroidal algebra, which we discuss later in Section \ref{sec:4}.
In the degenerate limit,
\begin{align}
    \Lambda_i(z)= 1+ \epsilon \frac{h_1h_2}{h_3}\sum_{r\in\mathbb{Z}} a_{-r}^{(i)} z^r +\mathcal{O}(\epsilon^2).\label{eq:Y00N_Lambda_deg}
\end{align}
It will be useful to recombine free bosons to include the off-diagonal parts to make the definition of the vertex operator simpler. For $r>0$, we define $h_r^{(i)}\;(i=1\sim N)$ as:
\begin{align}
    h_{-r}^{(i)}&=\frac{\kappa_r}{r} q_3^{-\frac{r}{2}N}
    \frac{q_3^{\frac{r}{2}(i-1)}}{(q_3^\frac{r}{2}-q_3^{-\frac{r}{2}})^2}
    a_{-r}^{(i)} , \notag\\
    h_0^{(i)}&=\epsilon\frac{h_1h_2}{h_3} a_0^{(i)} ,\notag\\
    h_r^{(i)}&=\frac{\kappa_r}{r} q_3^{\frac{r}{2}N} \frac{q_3^{-\frac{r}{2}i}}{(q_3^\frac{r}{2}-q_3^{-\frac{r}{2}})^2}
    a_r^{(i)}
    -\frac{\kappa_r}{r}q_3^{\frac{r}{2}N}
    \sum_{j=i+1}^N \frac{q_3^{-\frac{r}{2}(j-1)}}{q_3^\frac{r}{2}-q_3^{-\frac{r}{2}}}
    a_r^{(j)}.
\end{align}
The commutation relation between them is
\begin{align}
    [h_{-r}^{(i)}, h_r^{(j)}]=\frac{\kappa_r}{r}\frac{1}{q_3^r-1}\delta_{i,j}-\frac{\kappa_r}{r}\theta(i>j),
\end{align}
where $r$ is positive.
Using these bosons $h_{-r}^{(i)}$, we can rewrite the vertex operator $\Lambda_i$ as
\begin{align}
    \Lambda_i(z)=: \exp\left(\sum_{r\in\mathbb{Z}}h_{-r}^{(i)} z^r\right):,\quad (i=1,\cdots, N).
\end{align}
Using the vertex operator $\Lambda_i$, we define the quantum Miura operator as
\begin{align}
    R_i(z)=1-\Lambda_i(z) q_3^{-D_z}, \quad i=1,\cdots, N.
\end{align}
Here the derivative $D_z$ is defined as $D_z=z\frac{d}{dz}$ and the shift operator $q_3^{D_z}$ acts as
\begin{align}\label{eq:shiftop}
    q_3^{D_z}f(z)=f(q_3 z).
\end{align}
In the degenerate limit, by using (\ref{eq:Y00N_Lambda_deg}),
\begin{align}
    R_i(z)=\epsilon h_3\left(D_z-\frac{h_1h_2}{h_3^2}\sum_{r\in\mathbb{Z}} a_{-r}^{(i)} z^r\right)+\mathcal{O}(\epsilon^2) \equiv\epsilon h_3 z(\partial+\mu_3\partial\phi(z))+\mathcal{O}(\epsilon^2),\label{eq:Y00N_Ri_miura_multiplied_z}
\end{align}
which coincides with the Miura operator in Section \ref{sec:2} (\ref{eq:degenerate_Miuraop}) up to the multiplication of  $z$.

The quantum Miura transformation is constructed from the product of $N$ Miura operators as
\begin{align}
    : R_1(z) R_2(z)\cdots R_N(z) : =\sum_{i=0}^N (-1)^i T_i(z) q_3^{-i D_z}.
\end{align}
In the right-hand side, the coefficient of each order of $q_3$ is the generator of the algebra $Y_{0,0,N}$:
\begin{align}
    T_1(z) &= \Lambda_1(z)+\Lambda_2(z)+\cdots \Lambda_N(z),\label{eq:Y00N_T1}\\
    T_2(z) &= : \Lambda_1(z) \Lambda_2(q_3^{-1}z): +:\Lambda_1(z) \Lambda_3(q_3^{-1}z):
    +\cdots +: \Lambda_{N-1}(z) \Lambda_N(q_3^{-1}z) :,\\
    \vdots\notag\\
    T_N(z) &= :\Lambda_1(z)\Lambda_2(q_3^{-1}z)\cdots \Lambda_N(q_3^{-N+1}z):\label{eq:Y00N_TN}.
\end{align}

\subsubsection{Miura transformation for \texorpdfstring{$q$}{q}-\texorpdfstring{$W_N$}{WN}}
$W_N$ algebra can be obtained from $Y_{0,0,N}$ by subtracting $U(1)$ current. The procedure is parallel to the undeformed case.

We remove the diagonal $U(1)$ current from the the vertex operators $\Lambda_i$ to define $\tilde{\Lambda}_i\;(i=1\sim N)$,
\begin{align}
    &\quad\tilde{\Lambda}_i(z)\notag\\
    &=\exp\left[\epsilon\frac{h_1h_2}{h_3}\left(a_0^{(i)}-\frac{1}{N}\sum_{j=1}^N a_0^{(j)}\right)\right]
    \notag\\
    &\quad\times\exp\left[\sum_{r>0}-\frac{\kappa_r}{r}\frac{1}{q_3^{\frac{r}{2}N}-q_3^{-\frac{r}{2}N}}
    \sum_{j=1}^{N} \frac{q_3^{-\frac{r}{2}j}}{q_3^\frac{r}{2}-q_3^{-\frac{r}{2}}}
    a_{-r}^{(j)} z^r
    +\sum_{r>0}\frac{\kappa_r}{r} q_3^{-\frac{r}{2}N}
    \frac{q_3^{\frac{r}{2}(i-1)}}{(q_3^\frac{r}{2}-q_3^{-\frac{r}{2}})^2}
    a_{-r}^{(i)} z^r\right]\notag\\
    &\quad\times \exp\left[\sum_{r>0}-\frac{\kappa_r}{r}\frac{1}{q_3^{\frac{r}{2}N}-q_3^{-\frac{r}{2}N}}
    \sum_{j=1}^{N} \frac{q_3^{-\frac{r}{2}(j-1)}}{q_3^\frac{r}{2}-q_3^{-\frac{r}{2}}}
    a_r^{(j)} z^{-r}
    +\sum_{r>0}\frac{\kappa_r}{r} q_3^{\frac{r}{2}N} \frac{q_3^{-\frac{r}{2}i}}{(q_3^\frac{r}{2}-q_3^{-\frac{r}{2}})^2}
    a_r^{(i)} z^{-r}\right.\notag\\
    &\quad\quad\quad\left.+\sum_{r>0}-\frac{\kappa_r}{r}q_3^{\frac{r}{2}N}
    \sum_{j=i+1}^N \frac{q_3^{-\frac{r}{2}(j-1)}}{q_3^\frac{r}{2}-q_3^{-\frac{r}{2}}}
    a_r^{(j)} z^{-r}
    \right].
   \label{eq:Y00N_Lambda_tilde}
\end{align}
In the degenerate limit, the vertex operator reduces to
\begin{align}
    \tilde{\Lambda}_i(z)=1+\epsilon\frac{h_1h_2}{h_3}\sum_{r\in\mathbb{Z}}
    \left(a_{-r}^{(i)}-\frac{1}{N}\sum_{j=1}^N a_{-r}^{(j)}\right)z^r+\mathcal{O}(\epsilon^2)\label{eq:tildeLambda_deg}
\end{align}
This is the subtraction of the diagonal $U(1)$ which we met in section 2.

As in $Y_{0,0,N}$, we introduce free bosons $\tilde{h}_r^{(i)}\;(i=1\sim N)$ by combining the off-diagonal part. For $r>0$,
\begin{align}
    \tilde{h}_{-r}^{(i)}&=
    -\frac{\kappa_r}{r}\frac{1}{q_3^{\frac{r}{2}N}-q_3^{-\frac{r}{2}N}}
    \sum_{j=1}^N \frac{q_3^{-\frac{r}{2}j}}{q_3^\frac{r}{2}-q_3^{-\frac{r}{2}}}
    a_{-r}^{(j)}
    +\frac{\kappa_r}{r} q_3^{-\frac{r}{2}N}
    \frac{q_3^{\frac{r}{2}(i-1)}}{(q_3^\frac{r}{2}-q_3^{-\frac{r}{2}})^2}
    a_{-r}^{(i)},\notag\\
    \tilde{h}_0^{(i)}&=\epsilon\frac{h_1h_2}{h_3} \left(a_0^{(i)}-\frac{1}{N}\sum_{j=1}^N a_0^{(j)}\right),\notag\\
    \tilde{h}_r^{(i)}&=
    -\frac{\kappa_r}{r}\frac{1}{q_3^{\frac{r}{2}N}-q_3^{-\frac{r}{2}N}}
    \sum_{j=1}^N \frac{q_3^{-\frac{r}{2}(j-1)}}{q_3^\frac{r}{2}-q_3^{-\frac{r}{2}}}
    a_r^{(j)}
    +\frac{\kappa_r}{r} q_3^{\frac{r}{2}N} \frac{q_3^{-\frac{r}{2}i}}{(q_3^\frac{r}{2}-q_3^{-\frac{r}{2}})^2}
    a_r^{(i)}
    -\frac{\kappa_r}{r}q_3^{\frac{r}{2}N}
    \sum_{j=i+1}^N \frac{q_3^{-\frac{r}{2}(j-1)}}{q_3^\frac{r}{2}-q_3^{-\frac{r}{2}}}
    a_r^{(j)}.
\end{align}
The commutation relation is,
\begin{align}
    [\tilde{h}_{-r}^{(i)}, \tilde{h}_r^{(j)}]
    =\frac{\kappa_r}{r}\left(
    \frac{1}{q_3^{rN}-1}+\frac{1}{q_3^r-1}\delta_{i,j}-\theta(i>j)
    \right)
\end{align}
where $r$ is positive.
Due to the subtraction of the diagonal $U(1)$, $\tilde{h}_r^{(i)}$ satisfies the condition,
\begin{align}\label{eq:cond_tilde_hi}
    \sum_{i=1}^N q_3^{r i}\tilde{h}_r^{(i)}=0.
\end{align}
Using these bosons $\tilde{h}_{-r}^{(i)}$, the vertex operators $\tilde{\Lambda}_i$ is simplified,
\begin{align}
    \tilde{\Lambda}_i(z) = : \exp\left(\sum_{r\in \mathbb{Z}} \tilde{h}_{-r}^{(i)} z^r\right) : .
\end{align}
We define Miura operators as
\begin{align}
    R_i(z) = 1-\tilde{\Lambda}_i(z) q_3^{-D_z}.
\end{align}
In the degenerate limit $\epsilon\to 0$,
\begin{align}
    R_i=\epsilon h_3\left(D_z-\frac{h_1h_2}{h_3^2}\sum_{r\in\mathbb{Z}}\left(a_{-r}^{(i)}-\frac{1}{N}\sum_{i=1}^N a_{-r}^{(j)}\right)z^r\right)+\mathcal{O}(\epsilon^2).
\end{align}
From the same procedure as (\ref{eq:Y00N_Ri_miura_multiplied_z}), we see that this is the same as the Miura operators of $W_N$ algebra before the $q$-deformation up to the multiplication by $z$.

The quantum Miura transformation is constructed by the product of $N$ Miura operators as
\begin{align}
    : R_1(z)R_2(z)\cdots R_N(z) : =\sum_{i=0}^N (-1)^i T_i(z) q_3^{-i D_z}.
\end{align}
In the right-hand side, the coefficient of each order of $q_3$ is the generator of the algebra.
By expanding the above equation, we obtain the generators as
\begin{align}
    T_1(z) &= \tilde{\Lambda}_1(z)+\tilde{\Lambda}_2(z)+\cdots \tilde{\Lambda}_N(z),\\
    T_2(z) &= : \tilde{\Lambda}_1(z) \tilde{\Lambda}_2(q_3^{-1}z): +:\tilde{\Lambda}_1(z) \tilde{\Lambda}_3(q_3^{-1}z):
    +\cdots +: \tilde{\Lambda}_{N-1}(z) \tilde{\Lambda}_N(q_3^{-1}z) :,\\
    \vdots\notag\\
    T_N(z) &= :\tilde{\Lambda}_1(z)\tilde{\Lambda}_2(q_3^{-1}z)\cdots \tilde{\Lambda}_N(q_3^{-N+1}z):=1.
\end{align}
These are almost the same as (\ref{eq:Y00N_T1})-(\ref{eq:Y00N_TN}), but vertex operators $\Lambda_i$ are replaced with $\tilde{\Lambda}_i$. The only difference is that the generator $T_N(z)$ becomes an identity operator as a consequence of the subtraction of $U(1)$.

\subsubsection{Quadratic relations}
In the $q$-deformed case, the commutation relations (or the operator product expansion) in 2D CFT is replaced by so-called $fTT$ (or quadratic) relations. We review it here for the simplest example, $q$-Virasoro algebra, along the line of \cite{Shiraishi:1995rp}.

In this case, we have only one non-trivial generator,
\begin{align}
    T(z)=\tilde{\Lambda}_1(z)+\tilde{\Lambda}_2(z).
\end{align}
It satisfies the following equation, which is referred to as the $fTT$ (or the quadratic) relation,
\begin{align}
    \quad f\left(\frac{w}{z}\right) T(z) T(w)
    -f\left(\frac{z}{w}\right) T(w) T(z)
    =\frac{(q_1^\frac{1}{2}-q_1^{-\frac{1}{2}})(q_2^\frac{1}{2}-q_2^{-\frac{1}{2}})}{q_3^\frac{1}{2}-q_3^{-\frac{1}{2}}}
    \left(\delta\left(\frac{q_3 w}{z}\right)-\delta\left(\frac{w}{q_3 z}\right)\right),
\end{align}
where $\delta(z)$ is the delta function defined as
\begin{align}
    \delta(z)=\sum_{m\in\mathbb{Z}} z^m.
\end{align}
$f(z)$ in the above equation is defined as
\begin{align}
    f(z)=\exp\left[\sum_{m=1}^\infty
    \frac{\kappa_m}{q_3^m-q_3^{-m}}z^m
    \right].
\end{align}
A similar quadratic relation holds for all $q$-$W_N$ algebras \cite{Awata:1995zk}.
The $fTT$ relation reduces to the commutation relation in the degenerate limit.

\paragraph{Degenerate limit:}
The commutation relation of $\tilde{h}_r^{(i)}$ becomes
\begin{align}
    [\tilde{h}_{-r}^{(i)}, \tilde{h}_r^{(j)}]
        =\frac{r \epsilon^2 h_1h_2}{2}(1+2\delta_{i,j})+\mathcal{O}(\epsilon^3).
\end{align}
We note that $\tilde{h}_r^{(i)}\sim\mathcal{O}(\epsilon)$.
In the small $\epsilon$ limit, $T(z)$ is expanded as
\begin{align}
    T(z) = 2+\sum_{r\in \mathbb{Z}} (\tilde{h}_{-r}^{(1)}+\tilde{h}_{-r}^{(2)}) z^r
    +\frac{1}{2}\sum_{r,s\in\mathbb{Z}}(\tilde{h}_{-r}^{(1)}\tilde{h}_{-s}^{(1)}+\tilde{h}_{-r}^{(2)}\tilde{h}_{-s}^{(2)})z^{r+s}
    +\mathcal{O}(\epsilon^3).
\end{align}
Here it looks like that $T(z)$ has $\mathcal{O}(\epsilon)$ term, but because we have the condition (\ref{eq:cond_tilde_hi}) the the lowest term except the constant term is $\mathcal{O}(\epsilon^2)$.
By writing $h_r^{(i)}$ using $a_r^{(i)}$, we can rewrite $T(z)$ as
\begin{align}
    T(z)=2+\epsilon^2h_1h_2\left[\frac{h_3}{\sqrt{h_1h_2}}\sum_{r\in\mathbb{Z}} r \tilde{a}_{-r}z^r
    +\sum_{r,s\in\mathbb{Z}} :\tilde{a}_{-r}\tilde{a}_{-s}:z^{r+s}\right]
    +\mathcal{O}(\epsilon^4),
\end{align}
where $\tilde{a}_r$ is defined as
\begin{align}
    \tilde{a}_r=\frac{\sqrt{h_1h_2}}{h_3}\left(-\frac{1}{2}a_r^{(1)}+\frac{1}{2}a_r^{(2)}\right)
\end{align}
and it has the following commutation relation:
\begin{align}
    [\tilde{a}_r,\tilde{a}_s]=-\frac{r}{2}\delta_{r+s,0}.
\end{align}
One can obtain the Virasoro generator multiplied by $z^2$ from the coefficient of $-\epsilon^2 h_1h_2$ by subtracting some constants \cite{Shiraishi:1995rp},
\begin{align}
    L(z)=-\frac{h_3}{\sqrt{h_1h_2}}\sum_{r\in\mathbb{Z}} r\tilde{a}_{-r}z^{r-2}
    -\sum_{r,s\in\mathbb{Z}} :\tilde{a}_{-r}\tilde{a}_{-s}:z^{r+s-2}+\mathrm{const.},
\end{align}
so we see that $q$-Virasoro algebra becomes the Virasoro algebra in the degenerate limit.
%
\subsection{\texorpdfstring{$q$}{q}-deformation of the Miura operators from \texorpdfstring{$q$}{q}-Pochhammer}
To define the $q$-deformation of the Miura operators, we find that the $q$-analogue of the Pochhammer symbol is useful. For a comprehensive review, see for example \cite{Andrews1986}.

In the original Y-algebra, the quantum Miura transformation was represented by a fractional power differential operator (\ref{eq:degenerate_Miuraop}).
To illustrate the relevance of the Pochhammer symbol, we consider a simplified operator
$(\partial+\partial\phi)^\nu$, and examine a simplified situation where $\nu$ is a positive integer $n$.
We note that a formal expression in terms of Pochhammer symbol (\ref{Pochhammer}) gives a correspondence between the integer power differential operators.
We use $-D_z-D_z\phi=z(-\partial-\partial\phi)$ as the argument of Pochhammer symbol,
\begin{align}
    &\quad (-D_z-D_z\phi)_n\notag\\
    &=(z(-\partial-\partial\phi)+n-1)\cdots
    (z(-\partial-\partial\phi)+1)(z(-\partial-\partial\phi))\notag\\
    &=z^n (-\partial-\partial\phi)^n.
\end{align}
This is the Miura operator multiplied by $(-z)^n$.

An advantage of replacing the differential operator by the Pochhammer symbol is that the $q$-deformation ($q$-Pochhammer) is well-known:
\begin{align}
    (x;q)_n = \prod_{j=0}^{n-1}(1-x q^j).
\end{align}
In the limit of
\begin{align}
    x=q^y, q\to 1, y:\mathrm{fixed},
\end{align}
$q$-Pochhammer gives a proper definition of the $q$-deformation in the following sense,
\begin{align}
    \lim_{q\to 1} \frac{(x;q)_n}{(1-q)^n}=(y)_n .\label{eq:qpoch_poch}
\end{align}

One can also define the analytic continuation of the parameter $n$ to a complex number $\nu$, which is essential to rewrite the fractional power appearing in the Miura operator. For this purpose we introduce,
\begin{align}
	(x;q)_\infty =\prod_{j=0}^\infty (1-xq^j).
\end{align}
We can define the analytic continuation as,
\begin{align}
	(x;q)_\nu = \frac{(x;q)_\infty}{(q^\nu x; q)_\infty}.
\end{align}
One can derive that it reduces to the original definition when $\nu$ is an integer $n$,
\begin{align}
	\frac{(x;q)_\infty}{(q^n x;q)_\infty} = \frac{\prod_{j=0}^\infty (1-q^j x)}{\prod_{j=0}^\infty (1-q^{n+j}x)}
	= \prod_{j=0}^{n-1} (1-q^j x)
	= (x;q)_n.
\end{align}

Finally, we will use the $q$-binomial series to express the $q$-Pochhammer symbol in terms of the powers of the operator,
\begin{align}\label{eq:q_binom_thm}
    (x;q)_\nu = \sum_{k=0}^\infty (-x)^k q^{\frac{k(k-1)}{2}}
    \left[
    \begin{array}{c}
      \nu\\
      k
    \end{array}
  \right]_q,
\end{align}
where $\left[\begin{array}{c} \nu\\k\end{array}\right]_q$ is a $q$-binomial coefficient defined as
\begin{align}
    \left[\begin{array}{c} \nu\\k\end{array}\right]_q
    =\frac{(q;q)_\nu}{(q;q)_k (q;q)_{\nu-k}}.
\end{align}
When $\nu$ is a positive integer $n$, the summation over $k$ is truncated by $n$.
This is the $q$-deformation of the binomial theorem.
Otherwise, it defines the analytic continuation.

\subsection{Definition of quantum Miura transformation}
From the analysis of the previous subsection, we propose a $q$-deformed Miura transformation which gives the generators of  $q$-deformed $Y_{L,M,N}$. The main claims of the section are (\ref{qMiuraOp}) and (\ref{qMiura}). This proposal will be justified by,
\begin{itemize}
	\item It gives the generators of $q$-deformed $W(\mathfrak{sl}(2|1))$ \cite{Kojima2019} which corresponds to $Y_{0,1,2}$ after removing the $U(1)$ factor.
	\item The generators of $q$-deformed $Y_{L,M,N}$ can be naturally identified with a direct product representation of the quantum toroidal $\mathfrak{gl}_1$. 
	\item One can find the $q$-analog of the screening currents which commutes with the Miura transformation.
\end{itemize}
The first statement is provided in section \ref{Wsl21}. The second one is critical to justify our proposal since $q$-deformed $Y_{L,M,N}$ should be directly related to the quantum toroidal $\mathfrak{gl}_1$. Finally, the third statement follows the second observation. The commutativity with the screening currents becomes the consequence of the known result. We will explain them in section \ref{sec:4}.

\paragraph{Free bosons and vertex operators:}
We start by preparing the notation.
The Miura transformation before the $q$-deformation depends on $h_1,h_2,h_3$ and they satisfy $h_1+h_2+h_3=0$.
To construct the $q$-deformed version, we introduce $q_1, q_2, q_3$.
These satisfy $q_1q_2q_3=1$.
The relation between $q_c$ and $h_c$ is (\ref{eq:rel_qc_hc}).
We combine $q_1, q_2, q_3$ to define $q_{\vec{c}}$ as
\begin{align}
    q_{\vec{c}}=\prod_{i=1}^{L+M+N} q_{c_i}=q_1^L q_2^M q_3^N,
\end{align}
where $\vec{c}$ is defined as (\ref{eq:vecc}).
From (\ref{htoalpha}) and (\ref{eq:mu_nu_by_h}), $q_{\vec{c}}$ will be related to a central element of the toroidal $\mathfrak{gl}_1$, $C^\perp$, as we see in Section \ref{sec:4}.
As we will show later, the corresponding formula is (\ref{eq:toroidal_pit}) in terms of the quantum toroidal $\mathfrak{gl}_1$.

While the Miura transformation before the $q$-deformation is written by free bosons $J_i$ and derivatives, the $q$-deformed quantum Miura transformation can be written by the vertex operators $\Lambda_i$ and shift operators.

For the purpose of defining the vertex operators, we introduce the free bosons with the commutation relation:
\begin{align}
    [a_m^{(i)}, a_n^{(j)}]=m\frac{(q_{c_i}^\frac{m}{2}-q_{c_i}^{-\frac{m}{2}})^3}{-\kappa_m}\delta_{m+n,0}\delta_{i,j},\label{eq:q_boson}
\end{align}
which is a generalization of (\ref{eq:Y00N_q_boson}).
We check that $a^{(i)}$ is a $q$-deformation of the free boson in the degenerate limit (\ref{eq:Ji}).
The limit $q_c\to 1$ is the same as $\epsilon\to 0$.
The coefficient in the right-hand side of (\ref{eq:q_boson}) becomes
\begin{align}
    m\frac{(q_{c_i}^\frac{m}{2}-q_{c_i}^{-\frac{m}{2}})^3}{-\kappa_m}
    \to
    -m\frac{h_{c_i}^3}{\sigma}+\mathcal{O}(\epsilon),
\end{align}
in the limit $\epsilon\to 0$. This is the same as (\ref{eq:Ji}).

We define vertex operators $\Lambda_i$ by using free bosons $a_r^{(i)}$ as
\begin{align}
    \Lambda_i(z)
    &=\exp\left(\epsilon \frac{\sigma}{h_{c_i}^2} a_0^{(i)}\right)
    \exp\left[
    \sum_{r>0}\frac{\kappa_r}{r} q_{\vec{c}}^{-\frac{r}{2}}
    \frac{q_{c_1}^{\frac{r}{2}}\cdots q_{c_{i-1}}^{\frac{r}{2}}}{(q_{c_i}^\frac{r}{2}-q_{c_i}^{-\frac{r}{2}})^2}
    a_{-r}^{(i)} z^r\right]\notag\\
    &\quad\times \exp\left[
    \sum_{r>0}\frac{\kappa_r}{r} q_{\vec{c}}^\frac{r}{2} \frac{q_{c_1}^{-\frac{r}{2}}\cdots q_{c_i}^{-\frac{r}{2}}}{(q_{c_i}^\frac{r}{2}-q_{c_i}^{-\frac{r}{2}})^2}
    a_r^{(i)} z^{-r}
    +\sum_{r>0}-\frac{\kappa_r}{r}q_{\vec{c}}^\frac{r}{2}
    \sum_{j=i+1}^{L+M+N} \frac{q_{c_1}^{-\frac{r}{2}}\cdots q_{c_{j-1}}^{-\frac{r}{2}}}{q_{c_j}^\frac{r}{2}-q_{c_j}^{-\frac{r}{2}}}
    a_r^{(j)} z^{-r}
    \right].\label{Lambda_i}
\end{align}
As in $q$-$Y_{0,0,N}$, there is a mixing in the second line, which is related to the coproduct of the toroidal algebra.
In the limit $\epsilon\to 0$, the vertex operator becomes
\begin{align}
    \Lambda_i(z)= 1+\epsilon \frac{\sigma}{h_{c_i}^2}\sum_{r\in\mathbb{Z}} a_{-r}^{(i)} z^r  +\mathcal{O}(\epsilon^2).\label{eq:YLMN_Lambda_deg}
\end{align}

To simplify the appearance of the vertex operator, we recombine free bosons to define $h_r^{(i)}$.
With a positive $r$,
\begin{align}
    h_{-r}^{(i)}&=\frac{\kappa_r}{r} q_{\vec{c}}^{-\frac{r}{2}}
    \frac{q_{c_1}^{\frac{r}{2}}\cdots q_{c_{i-1}}^{\frac{r}{2}}}{(q_{c_i}^\frac{r}{2}-q_{c_i}^{-\frac{r}{2}})^2}
    a_{-r}^{(i)} , \notag\\
    h_0^{(i)}&=\epsilon \frac{\sigma}{h_{c_i}^2} a_0^{(i)},\notag\\
    h_r^{(i)}&=\frac{\kappa_r}{r} q_{\vec{c}}^\frac{r}{2} \frac{q_{c_1}^{-\frac{r}{2}}\cdots q_{c_i}^{-\frac{r}{2}}}{(q_{c_i}^\frac{r}{2}-q_{c_i}^{-\frac{r}{2}})^2}
    a_r^{(i)}
    -\frac{\kappa_r}{r}q_{\vec{c}}^\frac{r}{2}
    \sum_{j=i+1}^{L+M+N} \frac{q_{c_1}^{-\frac{r}{2}}\cdots q_{c_{j-1}}^{-\frac{r}{2}}}{q_{c_j}^\frac{r}{2}-q_{c_j}^{-\frac{r}{2}}}
    a_r^{(j)}.
\end{align}
We note that $h_0^{(i)}$ commutes with all generators.
The commutation relation of free bosons $h_r^{(i)} \; (i=1,2,\cdots, L+M+N)$ is
\begin{align}
    [h_{-r}^{(i)},h_r^{(j)}]=\frac{\kappa_r}{r} \frac{q_{c_i}^{-\frac{r}{2}}}{q_{c_i}^\frac{r}{2}-q_{c_i}^{-\frac{r}{2}}} \delta_{i,j}
    -\frac{\kappa_r}{r}\theta(i>j),
\end{align}
where $r$ is positive and all the other commutation relations vanish.
The vertex operators $\Lambda_i$ can be expressed by $h_{-r}^{(i)}$ as
\begin{align}
    \Lambda_i(z)
    =:\exp\left[\sum_{r\in\mathbb{Z}} h_{-r}^{(i)} z^r\right]:.
\end{align}
The operator products of $\Lambda_i$ satisfies the following formula:
\begin{align}
    \Lambda_i(z)\Lambda_j(w)=\begin{dcases}
\exp\left(\sum_{r=1}^\infty\frac{\kappa_r}{r(1-q_{c_i}^r)}\biggl(\frac{w}{z}\biggr)^r\right):\Lambda_i(z)\Lambda_j(w):\quad(i=j)\\
\ \ :\Lambda_i(z)\Lambda_j(w):\quad(i>j)\\
\exp\left(\sum_{r=1}^\infty\frac{\kappa_r}{r}\biggl(\frac{w}{z}\biggr)^r\right):\Lambda_i(z)\Lambda_j(w):\quad(i<j)\ \ .
\end{dcases}
\end{align}

\paragraph{Miura operator: } We are ready to describe the definition of the Miura operators:

\begin{screen}
Using the vertex operators $\Lambda_i$ and the $q$-deformed Pochhammer symbol, we define a $q$-deformation of the Miura operator as
\begin{align}
    R^{(c)}(z)&=:\left(q_3^\frac{1-\nu_c}{2}\Lambda(z)q_3^{-D_z}; q_3\right)_{\nu_c}:\notag\\
    &=\sum_{n=0}^\infty :\prod_{j=1}^n
    \left(-\frac{q_c^\frac{1}{2} q_3^{-\frac{j-1}{2}}-q_c^{-\frac{1}{2}} q_3^{\frac{j-1}{2}}}{q_3^\frac{j}{2}-q_3^{-\frac{j}{2}}}
    \Lambda(q_3^{-j+1}z)\right): q_3^{-nD_z}.\label{qMiuraOp}
\end{align}
\end{screen}
For each subscript $c$, more explicit formulae are
\begin{align}
    R^{(1)}(z)&=\sum_{n=0}^\infty :\prod_{j=1}^n
    \left(-\frac{q_1^\frac{1}{2} q_3^{-\frac{j-1}{2}}-q_1^{-\frac{1}{2}} q_3^{\frac{j-1}{2}}}{q_3^\frac{j}{2}-q_3^{-\frac{j}{2}}}
    \Lambda(q_3^{-j+1}z)\;\right): q_3^{-nD_z},\label{eq:MiuraOPdef1}\\
    R^{(2)}(z)&=\sum_{n=0}^\infty :\prod_{j=1}^n
    \left(-\frac{q_2^\frac{1}{2} q_3^{-\frac{j-1}{2}}-q_2^{-\frac{1}{2}} q_3^{\frac{j-1}{2}}}{q_3^\frac{j}{2}-q_3^{-\frac{j}{2}}}
    \Lambda(q_3^{-j+1}z)\right): q_3^{-nD_z},\label{eq:MiuraOPdef2}\\
    R^{(3)}(z)&=1 - \Lambda(z) q_3^{-D_z}. \label{eq:MiuraOPdef3}
\end{align}
We note that $R^{(3)}$ consists of finite terms, while $R^{(1)}$ and $R^{(2)}$ have an infinite number of terms.
This is parallel to the degenerate limit (\ref{eq:degenerate_Miuraop}), where $R^{(3)}$ is a differential operator of order one, while $R^{(1)}$ and $R^{(2)}$ are the pseudo-differential operators which are rewritten as infinite series of the differential operators.
From this analogy, we sometimes call $R^{(3)}$ a difference operator and call $R^{(1)}$ and $R^{(2)}$ pseudo-difference operators in the $q$-deformed case.

We may replace the parameter $q_3$ to $q_1$ or $q_2$ in (\ref{qMiuraOp}). With such choices the Miura operator $R^{(1)}$ or $R^{(2)}$ becomes finite instead of $R^{(3)}$.
This corresponds to an operation which interchages $R^{(c)}$.

\paragraph{Degenerate limit:}
To explain that it describes a $q$-deformation of the (pseudo-)differential operator, we use the relations in the previous subsection.
The parameters $\nu_1,\nu_2,\nu_3$ in (\ref{qMiuraOp}) are defined as
\begin{align}
	\nu_c &= \log_{q_3} q_c,
\end{align}
which are identical to (\ref{eq:mu_nu_by_h}) in the degenerate limit.
$D_z=z\frac{d}{dz}$ and a shift operator $q_3^{D_z}$ acts as (\ref{eq:shiftop}).
We also used the $q$-binomial series (\ref{eq:q_binom_thm}) to obtain the second line of (\ref{qMiuraOp}) from the first.

To derive the degenerate limit (\ref{eq:degenerate_Miuraop}) from (\ref{qMiuraOp}), we apply the relation between the Pochhammer and the $q$-deformed one (\ref{eq:qpoch_poch}), to the definition of the Miura operator $R^{(i)}$(\ref{qMiuraOp}). In this case, the deformation parameter is $q_3$.
$y$ in (\ref{eq:qpoch_poch}) is
\begin{align}
    y=-D_z+\frac{1-\nu_i}{2}
    +\frac{h_1h_2}{h_{c_i}^2}\sum_{r\in\mathbb{Z}}a_{-r}^{(i)}z^r,\label{eq:YLMN_y_def}
\end{align}
which rewrite
\begin{align}
    y=z(-\partial-\mu_{c_i}\partial \phi_i(z)).
\end{align}
When $n\in\mathbb{Z}$, the Pochhammer symbol $(y)_n$ is rewritten as
\begin{align}
    (y)_n=(y+n-1)\cdots(y+1)y=(-z)^n(\partial+\mu_{c_i}\partial\phi_i)^n.
\end{align}
By taking analytic continuation from $n\in \mathbb{Z}_{>0}$ to $\nu_{c_i}\in\mathbb{C}$, one obtains
\begin{align}
    (y)_{\nu_{c_i}}=(-z)^{\nu_{c_i}}(\partial+\mu_{c_i}\partial\phi_i)^{\nu_{c_i}}.\label{eq:YLMN_y_Miura}
\end{align}
This is the same as the Miura operator in the degenerate limit (\ref{eq:MiuraOp}) up to the first factor $(-z)^{\nu_{c_i}}$. When we multiply such operators in the Miura transformation, we move these factors to the left or right of the Miura transformation. It causes the shift of the zero mode of $\phi$, which can be absorbed in the redefinition.

\paragraph{Miura transformation: } As in the degenerate case (\ref{Miura}), the Miura transformation is defined as an ordered product of the Miura operators,
\begin{screen}
We mutiply the Miura operators $L$ $R^{(1)}$, $M$ $R^{(2)}$ and $N$ $R^{(3)}$, 
\begin{align}
    : R_1^{(c_1)}(z) R_2^{(c_2)}(z) \cdots R_{L+M+N}^{(c_{L+M+N})}(z) :
    =\sum_{n=0}^\infty (-1)^n T_n(z) q_3^{-nD_z}.\label{qMiura}
\end{align}
The expansion in terms of the shift operator $q_3^{-D_z}$ defines $T_1, T_2, T_3,\cdots$ which are the generators of the $q$-deformed $Y_{L, M, N}$.
\end{screen}
There are $(L+M+N)!/L!M!N!$ ordering of $R^{(1)}, R^{(2)}, R^{(3)}$.
As in the degenerate case, we expect that the algebra generated by $T_i$ are mutually isomorphic for the different choices of the ordering. We can prove this statement in the next section.

The explicit form of $T_i$ is
\begin{align}
    T_i(z) = \sum_{\substack{n_1+ \cdots\\+n_{L+M+N}=i}} \prod_{k=1}^{L+M+N} \left(
    \frac{q_{c_k}^\frac{1}{2} q_3^{\frac{1}{2}(1-j_k)} - q_{c_k}^{-\frac{1}{2}} q_3^{-\frac{1}{2}(1-j_k)}}{q_3^\frac{j_k}{2}-q_3^{-\frac{j_k}{2}}}\right)
    : \prod_{l=1}^{L+M+N} \left(
    \prod_{j_l=1}^{n_l} \Lambda_l(q_3^{-(\sum_{m=1}^{l-1}n_m +j_l-1)}z)\right) : .
\end{align}
The corresponding formula in the degenerate case is (\ref{Us}). Since the shift operator is easier to handle than the higher order differentiation, the $q$-deformed version is easier to study. Indeed, one can directly compare it with the generators of the quantum toroidal algebra in the next section.

\paragraph{Removal of the $U(1)$ factor:}
So far, we define $Y_{L,M,N}$ which contains the diagonal $U(1)$ current.
To compare our construction with the references of $q$-W algebras, such as \cite{Awata:1995zk} and \cite{Kojima2019}, we have to remove the $U(1)$ factor and we refer to the algebra thus obtained as $\tilde{Y}_{L,M,N}$.

What we need to do is parallel to the $q$-$W_N$ algebra. We redefine the vertex operator,
\begin{align}
    &\quad\tilde{\Lambda}_i(z)\notag\\
    &=\exp\left[\epsilon \frac{\sigma}{h_{c_i}^2} a_0^{(i)}
    -\epsilon\frac{\sigma}{Lh_1+Mh_2+Nh_3}\sum_{j=1}^{L+M+N}\frac{a_0^{(j)}}{h_{c_j}}\right]\notag\\
    &\quad\times\exp\left[\sum_{r>0}-\frac{\kappa_r}{r}\frac{1}{q_{\vec{c}}^\frac{r}{2}-q_{\vec{c}}^{-\frac{r}{2}}}
    \sum_{j=1}^{L+M+N} \frac{q_{c_1}^{-\frac{r}{2}}\cdots q_{c_j}^{-\frac{r}{2}}}{q_{c_j}^\frac{r}{2}-q_{c_j}^{-\frac{r}{2}}}
    a_{-r}^{(j)} z^r
    +\sum_{r>0}\frac{\kappa_r}{r} q_{\vec{c}}^{-\frac{r}{2}}
    \frac{q_{c_1}^{\frac{r}{2}}\cdots q_{c_{i-1}}^{\frac{r}{2}}}{(q_{c_i}^\frac{r}{2}-q_{c_i}^{-\frac{r}{2}})^2}
    a_{-r}^{(i)} z^r\right]\notag\\
    &\quad\times \exp\left[\sum_{r>0}-\frac{\kappa_r}{r}\frac{1}{q_{\vec{c}}^\frac{r}{2}-q_{\vec{c}}^{-\frac{r}{2}}}
    \sum_{j=1}^{L+M+N} \frac{q_{c_1}^{-\frac{r}{2}}\cdots q_{c_{j-1}}^{-\frac{r}{2}}}{q_{c_j}^\frac{r}{2}-q_{c_j}^{-\frac{r}{2}}}
    a_r^{(j)} z^{-r}
    +\sum_{r>0}\frac{\kappa_r}{r} q_{\vec{c}}^\frac{r}{2} \frac{q_{c_1}^{-\frac{r}{2}}\cdots q_{c_i}^{-\frac{r}{2}}}{(q_{c_i}^\frac{r}{2}-q_{c_i}^{-\frac{r}{2}})^2}
    a_r^{(i)} z^{-r}\right.\notag\\
    &\quad\quad\quad\left.+\sum_{r>0}-\frac{\kappa_r}{r}q_{\vec{c}}^\frac{r}{2}
    \sum_{j=i+1}^{L+M+N} \frac{q_{c_1}^{-\frac{r}{2}}\cdots q_{c_{j-1}}^{-\frac{r}{2}}}{q_{c_j}^\frac{r}{2}-q_{c_j}^{-\frac{r}{2}}}
    a_r^{(j)} z^{-r}
    \right].
   \label{eq:Lambda_tilde}
\end{align}
We note that our construction does not work for $L=M=N$ since
the factor $q_{\vec{c}}^\frac{r}{2}-q_{\vec{c}}^{-\frac{r}{2}}$ appearing in the denominator vanishes.
As in the degenerate case, $Y_{L,L,L}$ describes a trivial system, which we will see in the next section.
In the following, we ignore these singular cases.

In the limit $\epsilon\to 0$, the vertex operator becomes
\begin{align}
\tilde{\Lambda}_i(z)= 1+ \epsilon\sigma\sum_{r\in\mathbb{Z}}z^r\left(\frac{1}{h_{c_i}^2}
a_{-r}^{(i)}
-\frac{1}{Lh_1+Mh_2+Nh_3}\sum_{j=1}^{L+M+N}\frac{a_{-r}^{(j)}}{h_{c_j}}\right)+\mathcal{O}(\epsilon^2).
\end{align}
This is almost the same as (\ref{eq:YLMN_Lambda_deg}) but only the choice of the boson is different.

For convenience of calculation, we express free bosons as $\tilde{h}_r^{(i)}$. For $r>0$,
\begin{align}
    \tilde{h}_{-r}^{(i)}&=
    -\frac{\kappa_r}{r}\frac{1}{q_{\vec{c}}^\frac{r}{2}-q_{\vec{c}}^{-\frac{r}{2}}}
    \sum_{j=1}^{L+M+N} \frac{q_{c_1}^{-\frac{r}{2}}\cdots q_{c_j}^{-\frac{r}{2}}}{q_{c_j}^\frac{r}{2}-q_{c_j}^{-\frac{r}{2}}}
    a_{-r}^{(j)}
    +\frac{\kappa_r}{r} q_{\vec{c}}^{-\frac{r}{2}}
    \frac{q_{c_1}^{\frac{r}{2}}\cdots q_{c_{i-1}}^{\frac{r}{2}}}{(q_{c_i}^\frac{r}{2}-q_{c_i}^{-\frac{r}{2}})^2}
    a_{-r}^{(i)},\notag\\
    \tilde{h}_0^{(i)}&=\epsilon \frac{\sigma}{h_{c_i}^2} a_0^{(i)}
    -\epsilon\frac{\sigma}{Lh_1+Mh_2+Nh_3}\sum_{j=1}^{L+M+N}\frac{a_0^{(j)}}{h_{c_j}}
    ,\notag\\
    \tilde{h}_r^{(i)}&=
    -\frac{\kappa_r}{r}\frac{1}{q_{\vec{c}}^\frac{r}{2}-q_{\vec{c}}^{-\frac{r}{2}}}
    \sum_{j=1}^{L+M+N} \frac{q_{c_1}^{-\frac{r}{2}}\cdots q_{c_{j-1}}^{-\frac{r}{2}}}{q_{c_j}^\frac{r}{2}-q_{c_j}^{-\frac{r}{2}}}
    a_r^{(j)}
    +\frac{\kappa_r}{r} q_{\vec{c}}^\frac{r}{2} \frac{q_{c_1}^{-\frac{r}{2}}\cdots q_{c_i}^{-\frac{r}{2}}}{(q_{c_i}^\frac{r}{2}-q_{c_i}^{-\frac{r}{2}})^2}
    a_r^{(i)}
    -\frac{\kappa_r}{r}q_{\vec{c}}^\frac{r}{2}
    \sum_{j=i+1}^{L+M+N} \frac{q_{c_1}^{-\frac{r}{2}}\cdots q_{c_{j-1}}^{-\frac{r}{2}}}{q_{c_j}^\frac{r}{2}-q_{c_j}^{-\frac{r}{2}}}
    a_r^{(j)}.
\end{align}
We note that $\tilde{h}_0^{(i)}$ commutes with all terms.

The $L+M+N$ bosons $\tilde{h}_r^{(i)}\; (i=1\sim L+M+N)$ have the following commutation relation for a positive $r$,
\begin{align}
    &\quad [\tilde{h}_{-r}^{(i)},\tilde{h}_r^{(j)}] \notag\\
    &=\frac{\kappa_r}{r} \frac{1}{(q_{\vec{c}}^\frac{r}{2}-q_{\vec{c}}^{-\frac{r}{2}})^2}
    \sum_{k=1}^N q_{c_1}^{-r}\cdots q_{c_{k-1}}^{-r}(1-q_{c_k}^{-r})
    +\frac{\kappa_r}{r} \frac{q_{\vec{c}}^\frac{r}{2}}{q_{\vec{c}}^\frac{r}{2}-q_{\vec{c}}^{-\frac{r}{2}}}
    \sum_{k=j+1}^N q_{c_1}^{-r} \cdots q_{c_{k-1}}^{-r} (1-q_{c_k}^{-r})\notag\\
    &\quad-\frac{\kappa_r}{r} \frac{q_{\vec{c}}^\frac{r}{2}}{q_{\vec{c}}^\frac{r}{2}-q_{\vec{c}}^{-\frac{r}{2}}} q_{c_1}^{-r}\cdots q_{c_j}^{-r}
    -\frac{\kappa_r}{r}\frac{q_{\vec{c}}^{-\frac{r}{2}}}{q_{\vec{c}}^\frac{r}{2} - q_{\vec{c}}^{-\frac{r}{2}}}
    +\frac{\kappa_r}{r}\frac{q_{c_i}^{-\frac{r}{2}}}{q_{c_i}^\frac{r}{2}-q_{c_i}^{-\frac{r}{2}}} \delta_{i,j}
    -\frac{\kappa_r}{r}\theta(i>j).
\end{align}
All the other commutation relations vanish.

For $q$-$W_N$ algebra, the removal of the $U(1)$ factor is described by (\ref{eq:cond_tilde_hi}). For $\tilde{Y}_{L,M,N}$, the generalized formula is,
\begin{align}
    \sum_{i=1}^{L+M+N}(q_{c_i}^\frac{r}{2}-q_{c_i}^{-\frac{r}{2}}) q_{c_1}^r \cdots q_{c_{i-1}}^r q_{c_i}^\frac{r}{2} \tilde h_r^{(i)}=0.
\end{align}
Using $\tilde{h}_{-r}^{(i)}$, the vertex operator $\tilde{\Lambda}_i$ can be written as
\begin{align}
    \tilde{\Lambda}_i(z)
    =:\exp\left[\sum_{r\in\mathbb{Z}} \tilde{h}_{-r}^{(i)} z^r\right]:.
\end{align}
The normal ordering of the $\tilde{\Lambda}_i$ is
\begin{align}\label{eq:tildeLambda_no}
    \tilde{\Lambda}_i(z)\tilde{\Lambda}_j(w)=\begin{dcases}
\exp\left(\sum_{r=1}^\infty\frac{\kappa_r(q_{c_i}^r-q_{\vec{c}}^r)}{r(1-q_{c_i}^r)(1-q_{\vec{c}}^r)}\biggl(\frac{w}{z}\biggr)^r\right):\tilde{\Lambda}_i(z)\tilde{\Lambda}_j(w):\quad(i=j)\\
\exp\left(\sum_{r=1}^\infty\frac{-\kappa_r}{r(1-q_{\vec{c}}^r)}\biggl(\frac{w}{z}\biggr)^r\right):\tilde{\Lambda}_i(z)\tilde{\Lambda}_j(w):\quad(i>j)\\
\exp\left(\sum_{r=1}^\infty\frac{\kappa_r}{r(1-q_{\vec{c}}^{-r})}\biggl(\frac{w}{z}\biggr)^r\right):\tilde{\Lambda}_i(z)\tilde{\Lambda}_j(w):\quad(i<j)\quad.
\end{dcases}
\end{align}

We can obtain the Miura operator $R^{(i)}$ by replacing $\Lambda_i$ with $\tilde{\Lambda}_i$ in (\ref{qMiuraOp}).
The degenerate limit of $R^{(i)}$ is almost the same as before.
We consider the procedure from (\ref{eq:YLMN_y_def}) to (\ref{eq:YLMN_y_Miura}) and replace $y$ slightly as
\begin{align}
    y=-D_z+\frac{1-\nu_i}{2}+
    \left(\frac{h_1h_2}{h_{c_i}^2}
\sum_{r\in\mathbb{Z}}a_{-r}^{(i)}
-\frac{h_1h_2}{Lh_1+Mh_2+Nh_3}\sum_{r\in\mathbb{Z}}\sum_{j=1}^{L+M+N}\frac{a_{-r}^{(j)}}{h_{c_j}}\right)z^r.
\end{align}
Also in the case without $U(1)$ current, the $q$-deformed Miura operator reduces to the Miura operator in Section \ref{sec:2} in the degenerate limit.

\paragraph{$fTT$ relation for $\tilde{Y}_{L,M,N}$:}
When $i \leq j$, we define $f_{i,j}$ as
\begin{align}\label{eq:def_fij}
    f_{i,j}(z)=\exp\left[ \sum_{m=1}^\infty \frac{1}{m}
    (q_3^{\frac{i}{2}m} - q_3^{-\frac{i}{2}m})
    (q_{\vec{c}}^\frac{m}{2}q_3^{-\frac{j}{2}m}-q_{\vec{c}}^{-\frac{m}{2}}q_3^{\frac{j}{2}m})
    \frac{(q_1^\frac{m}{2}-q_1^{-\frac{m}{2}})(q_2^\frac{m}{2}-q_2^{-\frac{m}{2}})}{(q_{\vec{c}}^\frac{m}{2}-q_{\vec{c}}^{-\frac{m}{2}})(q_3^\frac{m}{2}-q_3^{-\frac{m}{2}})}z^m
    \right].
\end{align}
We note that the structure functions only depend on $\vec c$ through $q_{\vec{c}}$. This implies that the quadratic relations does not depend on the color order $\vec c$, and the algebra of $q$-$\tilde{Y}_{L,M,N}$ is unique. For $i\geq j$, we define $f_{i,j}(z)=f_{j,i}(z)$.


We conjecture that the quadratic relation between $T$s is as follows,
\begin{screen}
\begin{align}\label{eq:fTT_pred}
    &\quad f_{i,j}\left(\frac{q_3^\frac{i-j}{2} w}{z}\right) T_i(z) T_j(w)
    - f_{j,i}\left( \frac{q_3^\frac{j-i}{2}z}{w}\right) T_j(w) T_i(z)\notag\\
    &= \frac{(q_1^\frac{1}{2} - q_1^{-\frac{1}{2}})(q_2^\frac{1}{2}-q_2^{-\frac{1}{2}})}{q_3^\frac{1}{2} -q_3^{-\frac{1}{2}}}
    \sum_{k=1}^i \prod_{l=1}^{k-1} \frac{(1-q_1 q_3^{-l})(1-q_2q_3^{-l})}{(1-q_3^{-l-1})(1-q_3^{-l})}\notag\\
    &\times\left(
    \delta\left(\frac{q_3^k w}{z}\right) f_{i-k,j+k}(q_3^\frac{i-j}{2})
    T_{i-k}(q_3^{-k}z) T_{j+k}(q_3^k w)
    - \delta\left(\frac{q_3^{i-j-k}w}{z}\right) f_{i-k,j+k}(q_3^\frac{j-i}{2}) T_{i-k}(z) T_{j+k}(w)\right),
\end{align}
where we assume $i\leq j$.
\end{screen}
The relation is a generalization of $fTT$ relation in \cite{Kojima2019}, where the author gives a description of $W(\mathfrak{sl}(2|1))$ algebra, which corresponds to $\tilde{Y}_{0,1,2}$. 
We give a partial proof of the quadratic relations when one of the generators is the lowest generator $T_{1}(z)$ in the Appendix. The quadratic relations for other currents can be obtained by using the fusion formulas as in the main theorem in \cite{Kojima2019}.
\subsection{\texorpdfstring{$(L, M, N)=(0, 1, 2)$}{(L,M,N)=(0,1,2)} case: \texorpdfstring{$W_{q,t}(\mathfrak{sl}(2|1))$}{Wq,t(sl(2|1))}}\label{Wsl21}
In this section, we show that  the $q$-deformed $\tilde{Y}_{0,1,2}$ is equivalent to  $W_{q,t}(\mathfrak{sl}(2|1))$ proposed in  \cite{Ding1999}. The $q$-W algebra $W_{q,t}(\mathfrak{sl}(2|1))$ is defined by the screening charges associated with $\mathfrak{sl}(2|1)$ and realized by two free bosons. Because there are two types of Dynkin diagrams for $\mathfrak{sl}(2|1)$, $W_{q,t}(\mathfrak{sl}(2|1))$ has two free boson representations. As we discussed in Section \ref{sec:2}, that corresponds to the fact that the order of the Miura operators is not unique. As we have seen, the bosonic (resp. fermionic) screening charge arises between the neighboring Miura operators of the same (resp. different) type.

In the following, we compare  the generators derived from the Miura transformation with those for $W_{q,t}(\mathfrak{sl}(2|1))$, whose detail is studied recently in \cite{Kojima2019}. 
This result supports our proposal of the Miura transformation.

\subsubsection{One bosonic, one fermionic case: \texorpdfstring{$R^{(3)} R^{(3)} R^{(2)}$}{R(3)R(3)R(2)}}
In our approach, we first define the Miura operators $R^{(i)}$, and by using the Miura transformation we obtain the generators.
In this case, $(L,M,N)=(0,1,2)$, so we use one $R^{(2)}$ and two $R^{(3)}$.
The ordering of the $R^{(2)}$ and $R^{(3)}$ corresponds to the number of bosonic currents and fermionic currents.
Here we consider one bosonic and one fermionic case, so one screening current between the same type Miura operators and one screening current between the different type Miura operators are needed.
There is two choices, which satisfies the above condition, $R^{(3)}R^{(3)}R^{(2)}$ and $R^{(2)}R^{(3)}R^{(3)}$, which should be equivalent.
Here we use the ordering $R^{(3)}R^{(3)}R^{(2)}$, which is directly comparable to \cite{Kojima2019}.

The Miura operators (\ref{qMiuraOp}) are written as,
\begin{align}
    R_1^{(3)}(z)&=1 - \tilde{\Lambda}_1(z) q_3^{-D_z},\\
    R_2^{(3)}(z)&=1 - \tilde{\Lambda}_2(z) q_3^{-D_z},\\
    R_3^{(2)}(z)&=\sum_{n=0}^\infty :\prod_{j=1}^n
    \left(-\frac{q_2^\frac{1}{2} q_3^{-\frac{j-1}{2}}-q_2^{-\frac{1}{2}} q_3^{\frac{j-1}{2}}}{q_3^\frac{j}{2}-q_3^{-\frac{j}{2}}}
    \tilde{\Lambda}_3(q_3^{-j+1}z)\right): q_3^{-nD_z}.
\end{align}
We use $\tilde{\Lambda}(z)$ since we need to subtract the $U(1)$ factor.
The Miura transformation (\ref{qMiura}) becomes
\begin{align}
    : R_1^{(3)}(z) R_2^{(3)}(z) R_3^{(2)}(z) :
    = \sum_{n=0}^\infty (-1)^n T_n(z) q_3^{-nD_z}.
\end{align}
The coefficients of the above expansion become,
\begin{align}
    T_1(z)&=\tilde{\Lambda}_1(z)+\tilde{\Lambda}_2(z)+\frac{q_2^\frac{1}{2}-q_2^{-\frac{1}{2}}}{q_3^\frac{1}{2}-q_3^{-\frac{1}{2}}} \tilde{\Lambda}_3(z),\\
    T_2(z)&=:\tilde{\Lambda}_1(z)\tilde{\Lambda}_2(q_3^{-1} z):
    + \frac{q_2^\frac{1}{2}-q_2^{-\frac{1}{2}}}{q_3^\frac{1}{2}-q_3^{-\frac{1}{2}}} :\tilde{\Lambda}_1(z)\tilde{\Lambda}_3(q_3^{-1} z):\notag\\
    &\quad + \frac{q_2^\frac{1}{2}-q_2^{-\frac{1}{2}}}{q_3^\frac{1}{2}-q_3^{-\frac{1}{2}}} :\tilde{\Lambda}_2(z)\tilde{\Lambda}_3(q_3^{-1} z):
    + \frac{q_2^\frac{1}{2}-q_2^{-\frac{1}{2}}}{q_3^\frac{1}{2}-q_3^{-\frac{1}{2}}}
    \frac{q_2^\frac{1}{2}q_3^{-\frac{1}{2}}-q_2^{-\frac{1}{2}}q_3^\frac{1}{2}}{q_3-q_3^{-1}}
    :\tilde{\Lambda}_3(z)\tilde{\Lambda}_3(q_3^{-1} z):,\\
    T_{n\geq3}(z)&=:\prod_{j=1}^n \frac{q_2^\frac{1}{2} q_3^{-\frac{j-1}{2}}-q_2^{-\frac{1}{2}} q_3^{\frac{j-1}{2}}}{q_3^\frac{j}{2}-q_3^{-\frac{j}{2}}}\tilde{\Lambda}_3(q_3^{-j+1}z):
    +:\tilde{\Lambda}_1(z) \prod_{j=1}^{n-1}\frac{q_2^\frac{1}{2} q_3^{-\frac{j-1}{2}}-q_2^{-\frac{1}{2}} q_3^{\frac{j-1}{2}}}{q_3^\frac{j}{2}-q_3^{-\frac{j}{2}}}\tilde{\Lambda}_3(q_3^{-j}z): \notag\\
    &\quad +:\tilde{\Lambda}_2(z) \prod_{j=1}^{n-1}\frac{q_2^\frac{1}{2} q_3^{-\frac{j-1}{2}}-q_2^{-\frac{1}{2}} q_3^{\frac{j-1}{2}}}{q_3^\frac{j}{2}-q_3^{-\frac{j}{2}}}\tilde{\Lambda}_3(q_3^{-j}z):\notag\\
    &\quad+:\tilde{\Lambda}_1(z)\tilde{\Lambda}_2(q_3^{-1}z) \prod_{j=1}^{n-2}\frac{q_2^\frac{1}{2} q_3^{-\frac{j-1}{2}}-q_2^{-\frac{1}{2}} q_3^{\frac{j-1}{2}}}{q_3^\frac{j}{2}-q_3^{-\frac{j}{2}}}\tilde{\Lambda}_3(q_3^{-j-1}z):.
\end{align}
$f_{i,j}\; (i\leq j)$ is give by (\ref{eq:def_fij}),
\begin{align}
    &\quad f_{i, j}(z)\notag\\
    &=\exp\left(-\sum_{m=1}^\infty
    \frac{1}{m} (q_3^{\frac{i}{2}m} - q_3^{-\frac{i}{2}m})
    (q_2^\frac{m}{2} q_3^{\left(1-\frac{j}{2}\right)m} - q_2^{-\frac{m}{2}} q_3^{-\left(1-\frac{j}{2}\right)m})
    \frac{(q_1^\frac{m}{2}-q_1^{-\frac{m}{2}})(q_2^\frac{m}{2}-q_2^{-\frac{m}{2}})}{(q_2^\frac{m}{2}q_3^m - q_2^{-\frac{m}{2}}q_3^{-m})(q_3^\frac{m}{2}-q_3^{-\frac{m}{2}})}
    z^m\right),\label{eq:fij_012}
\end{align}
and the $fTT$ relation (\ref{eq:fTT_pred}) is satisfied.

If we change parameters as
\begin{align}
    q_1=x^{2r},\quad
    q_2=x^{-2(r-1)},\quad
    q_3=x^{-2},
\end{align}
the generators are the same as those in \cite{Kojima2019} up to difference of the argument $z$,
and we have the same $f_{i,j}$. Thus the quadratic relation coincides exactly if we shift the argument $z$ of $T_i(z)$ by $q_3^i$.

\subsubsection{Two fermionic case: \texorpdfstring{$R^{(3)} R^{(2)} R^{(3)}$}{R(3)R(2)R(3)}}
Next, we consider the two fermionic case.
Now $(L,M,N)=(0,1,2)$, we have one $R^{(2)}$ and two $R^{(3)}$.
The procedures are similar to the one bosonic and one fermionic case.
In this case, two screening currents between the different type Miura operators are needed.
The only ordering which satisfies the above condition is $R^{(3)} R^{(2)} R^{(3)}$.

By (\ref{qMiuraOp}), the Miura operators are written as
\begin{align}
    R_1^{(3)}(z)&=1 - \tilde{\Lambda}_1(z) q_3^{-D_z},\\
    R_2^{(2)}(z)&=\sum_{n=0}^\infty :\prod_{j=1}^n
    \left(-\frac{q_2^\frac{1}{2} q_3^{-\frac{j-1}{2}}-q_2^{-\frac{1}{2}} q_3^{\frac{j-1}{2}}}{q_3^\frac{j}{2}-q_3^{-\frac{j}{2}}}
    \tilde{\Lambda}_2(q_3^{-j+1}z)\right): q_3^{-nD_z},\\
    R_3^{(3)}(z)&=1 - \tilde{\Lambda}_3(z) q_3^{D_z},
\end{align}
The definition of the quantum Miura transformation (\ref{qMiura}) becomes
\begin{align}
    : R_1^{(3)}(z) R_2^{(2)}(z) R_3^{(3)}(z) :
    = \sum_{n=0}^\infty (-1)^n T_n(z) q_3^{-nD_z}.
\end{align}
We obtain the generators as,
\begin{align}
    T_1(z)&=\tilde{\Lambda}_1(z)
    +\frac{q_2^\frac{1}{2}-q_2^{-\frac{1}{2}}}{q_3^\frac{1}{2}-q_3^{-\frac{1}{2}}} \tilde{\Lambda}_2(z)
    +\tilde{\Lambda}_3(z),\\
    T_2(z)&=:\tilde{\Lambda}_1(z)\tilde{\Lambda}_3(q_3^{-1} z):
    + \frac{q_2^\frac{1}{2}-q_2^{-\frac{1}{2}}}{q_3^\frac{1}{2}-q_3^{-\frac{1}{2}}} :\tilde{\Lambda}_1(z)\tilde{\Lambda}_2(q_3^{-1} z):\notag\\
    &\quad + \frac{q_2^\frac{1}{2}-q_2^{-\frac{1}{2}}}{q_3^\frac{1}{2}-q_3^{-\frac{1}{2}}} :\tilde{\Lambda}_2(z)\tilde{\Lambda}_3(q_3^{-1} z):
    + \frac{q_2^\frac{1}{2}-q_2^{-\frac{1}{2}}}{q_3^\frac{1}{2}-q_3^{-\frac{1}{2}}}
    \frac{q_2^\frac{1}{2}q_3^{-\frac{1}{2}}-q_2^{-\frac{1}{2}}q_3^\frac{1}{2}}{q_3-q_3^{-1}}
    :\tilde{\Lambda}_2(z)\tilde{\Lambda}_2(q_3^{-1} z):,\\
    T_{n\geq3}(z)&=:\prod_{j=1}^n \frac{q_2^\frac{1}{2} q_3^{-\frac{j-1}{2}}-q_2^{-\frac{1}{2}} q_3^{\frac{j-1}{2}}}{q_3^\frac{j}{2}-q_3^{-\frac{j}{2}}}\tilde{\Lambda}_2(q_3^{-j+1}z):
    +:\tilde{\Lambda}_1(z) \prod_{j=1}^{n-1}\frac{q_2^\frac{1}{2} q_3^{-\frac{j-1}{2}}-q_2^{-\frac{1}{2}} q_3^{\frac{j-1}{2}}}{q_3^\frac{j}{2}-q_3^{-\frac{j}{2}}}\tilde{\Lambda}_2(q_3^{-j}z): \notag\\
    &\quad +:\prod_{j=1}^{n-1}\frac{q_2^\frac{1}{2} q_3^{-\frac{j-1}{2}}-q_2^{-\frac{1}{2}} q_3^{\frac{j-1}{2}}}{q_3^\frac{j}{2}-q_3^{-\frac{j}{2}}}\tilde{\Lambda}_2(q_3^{-j+1}z) \tilde{\Lambda}_3(q_3^{-n+1}z):\notag\\
    &\quad+:\tilde{\Lambda}_1(z) \prod_{j=1}^{n-2}\frac{q_2^\frac{1}{2} q_3^{-\frac{j-1}{2}}-q_2^{-\frac{1}{2}} q_3^{\frac{j-1}{2}}}{q_3^\frac{j}{2}-q_3^{-\frac{j}{2}}}\tilde{\Lambda}_2(q_3^{-j}z) \tilde{\Lambda}_3(q_3^{-n+1}z):.
\end{align}
$f_{i,j}\; (i\leq j)$ is the same as (\ref{eq:fij_012}),
and the $fTT$ relation (\ref{eq:fTT_pred}) is satisfied.
If we choose the parameters as
\begin{align}
    q_1=x^{2r},\quad
    q_2=x^{-2(r-1)},\quad
    q_3=x^{-2},
\end{align}
our results are consistent with \cite{Kojima2019}.

\section{Miura transformation from quantum toroidal \texorpdfstring{$\mathfrak{gl}_1$}{gl1}} \label{sec:4}
The quantum toroidal $\mathfrak{gl}_1$ is a $q$-deformation of $W_{1+\infty}$. It contains parameters $q_1, q_2, q_3$ and there is a Fock realization $\mathcal{F}_c$ ($c=1,2,3$) associated with each of them. The main claim of this section is that the Miura transformation in the previous section, namely, the definition of higher currents by taking a product of the Miura operators in an order, corresponds to taking a coproduct of these Fock spaces in the same order. For instance, the Miura transformation for the quantum $W_N$ algebra with the $U(1)$ current corresponds to the tensor product of $N$ Fock spaces of the same type.

In \cite{bershtein2018plane}, the authors studied the coproduct of the Fock spaces of two different types in detail. They found that the screening charges associated with the $\mathfrak{gl}_{n|m}$ root system characterizes the corresponding quantum W-algebras.

In this section, we prove the claim by expressing the generators, obtained from the Miura transformation, as the product of the Drinfeld currents of the quantum toroidal algebra, and show that they are identical.

\subsection{Quantum toroidal \texorpdfstring{$\mathfrak{gl}_1$}{gl1}}
The quantum toroidal  $\mathfrak{gl}_1$\ \footnote{We will use the notation in  \cite{bershtein2018plane} in this section.} denoted by  $\mathcal{E}_1(q_1,q_2,q_3)$ has parameters $q_c=e^{\epsilon h_c}\ (c=1,2,3) $ with the constraint $q_1q_2q_3=1$, and $q\equiv q_3^{1/2}$. 
It is generated by the Drinfeld currents,
\begin{align*}
	E(z)=\sum_{m \in \mathbb{Z}}  E_mz^{-m},\;\; F(z)=\sum_{m \in \mathbb{Z}}  F_mz^{-m},\;\;  K^\pm(z)=(C^\perp)^{\pm 1} \exp\left(\sum_{r>0} \mp\frac{\kappa_r}{ r} H_{\pm r} z^{\mp r} \right),
\end{align*}
with the central elements $C,C^{\perp}$ and two degree operators $D, D^\perp$. Here, we use $\kappa_r$ defined in (\ref{eq:kappa}).

The defining relation of the quantum toroidal $\mathfrak{gl}_1$ is given as follows:
\begin{align*}
	& DE(z)=E(qz)D,\quad DF(z)=F(qz)D,\quad DK^{\pm}(z)= K^{\pm}(qz)D,\\
	& D^\perp E(z)=qE(z)D^\perp,\quad D^\perp F(z)=q^{-1}F(z)D^\perp,\quad \left[D^\perp, K^{\pm}(z)\right]=0\,,\\
	&g(z,w)E(z)E(w)+g(w,z)E(w)E(z)=0, \qquad\quad g(w,z)F(z)F(w)+g(z,w)F(w)F(z)=0,\\
	&K^\pm(z)K^\pm(w) = K^\pm(w)K^\pm (z), 
	\quad \qquad
	\frac{g(C^{-1}z,w)}{g(C z,w)}K^-(z)K^+ (w) 
	=
	\frac{g(w,C^{-1}z)}{g(w,C z)}K^+(w)K^-(z),
	\\
	&g(z,w)K^\pm(C^{(-1\mp1)/2}z)E(w)
	+g(w,z)E(w)K^\pm(C^{(-1\mp1) /2}z)=0,
	\\
	&g(w,z)K^\pm(C^{(-1\pm1)/2}z)
	F(w)+g(z,w)F(w)K^\pm(C^{(-1\pm1)/2}z)=0\,,
	\\
	&[E(z),F(w)]=\frac{1}{\kappa_1}
	(\delta\bigl(\frac{Cw}{z}\bigr)K^+(w)
	-\delta\bigl(\frac{Cz}{w}\bigr)K^-(z)),\\
	&\mathop{\mathrm{Sym}}_{z_1,z_2,z_3}z_2z_3^{-1}
	[E(z_1),[E(z_2),E(z_3)]]=0,\qquad\quad \mathop{\mathrm{Sym}}_{z_1,z_2,z_3}z_2z_3^{-1}
	[F(z_1),[F(z_2),F(z_3)]]=0,
\end{align*}
where
\begin{align*}
	g(z,w)=\prod_{c=1}^3 (z-q_cw).
\end{align*}
The generator $H_r$ in the definition of $K^\pm$ satisfies the following commutation relations:
\bq
\begin{split}
	&[H_r,H_s]=\delta_{r+s,0}r\frac{C^r-C^{-r}}{\kappa_r},\\
	&[H_r,E(z)]=-C^{(-r-|r|)/2}E(z)z^r,\\
	&[H_r,F(z)]=C^{(-r+|r|)/2}F(z)z^r.
\end{split}
\eq
They are equivalent to some of the defining relations.

For later convenience, we introduce a current $t(z)$ which commutes with the Heisenberg subalgebra generated by $H_r$ as follows \cite{FHSSY:2010}: 
\bq
\begin{split}
	&t(z)=\alpha(z)E(z)\beta(z),\\
	&\alpha(z)=\mathrm{exp}\bigl(\sum_{r=1}^{\infty}\frac{-\kappa_r}{r(1-C^{2r})}H_{-r}z^r\bigr),\\
	&\beta(z)=\mathrm{exp}\bigl(\sum_{r=1}^{\infty}\frac{-C^{-r}\kappa_r}{r(1-C^{-2r})}H_rz^{-r}\bigr).
\end{split}
\eq

As we mentioned in Section \ref{sec:2}, the quantum toroidal $\mathfrak{gl}_1$ has a plane partition representation \cite{feigin2012quantum}, where one of the central elements is set to $C=1$.  While it is irreducible for generic value of $C^\perp$, there appears null states (``pit") at $(L,M,N)$ if we set,
\ba
C^\perp=q_1^{L/2}q_2^{M/2}q_3^{N/2}.
\label{eq:toroidal_pit}
\ea 
It corresponds to the condition in the affine Yangian (\ref{DegLMN}).
We note that $q_1 q_2 q_3=1$ implies that $C^\perp$ is invariant under the shift symmetry (\ref{shift}).
In particular, for $L=M=N$, the central charges becomes trivial $C=C^\perp=1$. It means that we have a pit at $(0,0,0)$ and there is no non-vanishing states except for the highest weight state.

We remark here that the quantum toroidal $\mathfrak{gl}_1$ is more symmetric than the affine Yangian. It is symmetric with respect to ``S-duality" transformation  referred to as Miki automorphism.
\begin{align}\label{eq:Mikiduality}
	& C^\perp\rightarrow C\rightarrow (C^\perp)^{-1} \rightarrow C^{-1}\\
	& D^\perp\rightarrow D\rightarrow (D^\perp)^{-1} \rightarrow D^{-1}	\\
	& E_0 \rightarrow H_{-1}\rightarrow F_0\rightarrow H_1
\end{align}
It maps the spectral parameter to the world sheet parameter, and the plane partition representation with a ``pit" to the free boson representation.  It allows us to describe the ($q$-deformed) $Y_{L,M,N}$ from totally different viewpoint. 

For instance, the most fundamental representation where the pit is located at $(1,0,0)$, $(0,1,0)$, $(0,0,1)$
are originally defined as $C=1$ and $C^\perp=q_c^{1/2}$ for $c=1,2,3$. Through the Miki automorphism, they are mapped to the Fock representation with \footnote{In general, the quantum toroidal algebra has $SL(2,\mathbb{Z})$ automorphism. While we can realize $(C,C^\perp)=(q_c^{1/2},q_c^{n/2})$ ($c=1,2,3$) representations by free bosons, the free field description for other cases is difficult. Therefore, the action of the general element of $SL(2,Z)$ will be hard to describe.},
\ba
C=q_c^{1/2},\quad C^\perp=1\quad(c=1,2,3).\label{eq:DIM_center}
\ea
We note that there are three kinds of representations, which can be shuffled by the triality.  
For each $c$, we denote the corresponding Fock module by $\mathcal{F}_c(u)$. We fix the normalization of the free boson in the same way as (\ref{eq:q_boson}):\footnote{For simplicity, we do not put the subscript $c$ on $a_r$.}
\begin{equation}
[a_r,a_s]=r\frac{(q_c^{r/2}-q_c^{-r/2})^3}{-\kappa_{r}}\delta_{r+s,0}.
\label{eq:ar}
\end{equation}
The Drinfeld currents are realized as the vertex operators: 
\bq
\rho_u^{(c)}(E(z))=\frac{1-q_c}{\kappa_1}\eta_c(z),\quad\rho_u^{(c)}(F(z))=\frac{1-q_c^{-1}}{\kappa_1}\xi_c(z), \quad
\rho_u^{(c)}(H_r)=\frac{a_r}{q_c^{r/2}-q_c^{-r/2}},
\eq
where
\ba
&&\eta_c(z)=u\exp\left(\sum_{r=1}^\infty\frac{q_c^{-r/2}\kappa_r}{r(q_c^{r/2}-q_c^{-r/2})^2}a_{-r}z^r\right)\exp\left(\sum_{r=1}^\infty\frac{\kappa_r}{r(q_c^{r/2}-q_c^{-r/2})^2}a_{r}z^{-r}\right),\\
&&\xi_c(z)=u^{-1}\exp\left(\sum_{r=1}^\infty\frac{-\kappa_r}{r(q_i^{r/2}-q_i^{-r/2})^2}a_{-r}z^r\right)\exp\left(\sum_{r=1}^\infty\frac{-q_i^{r/2}\kappa_r}{r(q_i^{r/2}-q_i^{-r/2})^2}a_{r}z^{-r}\right).
\ea
where $u$ is the normalization parameter. We also denote
\ba
\rho_u^{(c)}(K^\pm(z))=\varphi_c^\pm(z)=
\exp\left(\sum_{r=1}^{\infty}\frac{-\kappa_r}{r(q_c^{r/2}-q_c^{-r/2})}a_{\pm r}z^{\mp r}\right).
\ea

To obtain the representation with $C=q_1^{L/2}q_2^{M/2}q_3^{N/2}$ which is the counterpart of (\ref{eq:toroidal_pit}), we need to consider the tensor product of the Fock spaces.
The action of the Drinfeld currents are determined from the coproduct:
\begin{equation}
\begin{aligned}
&\Delta(H_r)=H_r\otimes 1+C^{-r}\otimes H_r,\quad 
\Delta(H_{-r})=H_{-r}\otimes C^{r}+1\otimes H_{-r}, \quad r>0 
\\
&\Delta(E(z))=E\left(C_2^{-1}z\right)\otimes K^+\left(C_2^{-1}z\right)+ 1\otimes E\left(z\right),\\
&\Delta(F(z))=F\left(z\right)\otimes 1 + K^-\left(C_1^{-1}z\right)\otimes F(C_1^{-1}z),\\
&\Delta(X)=X\otimes X,\;\; \text{for $X= C, C^\perp$},
\end{aligned} \label{eq:coprod}
\end{equation}
where $C_1 =C\otimes 1$, $C_2 =1\otimes C$.
For convenience, we set $L+M+N$ as $n$.
Because the coproduct of  the central element is defined multiplicatively, the representation which is of our interest is realized by the tensor product of $L$ pieces of $\mathcal{F}_1$, $M$ pieces of $\mathcal{F}_2$ and $N$ pieces of $\mathcal{F}_3$ in an arbitrary order $\vec{c}$ in (\ref{eq:vecc}). In the following, we write $\mathcal{F}_{\vec{c}}(\vec{u})\equiv\mathcal{F}_{c_1}\otimes\mathcal{F}_{c_2}\otimes\cdots\mathcal{F}_{c_n}$. 
As an example, we consider the action of $E(z)$ on $\mathcal{F}_{\vec{c}}(\vec{u})$. We set 
\bq
\begin{split}
&\Delta^{(m)}=(1\otimes\Delta)\Delta^{(m-1)},\quad\Delta^{(1)}=\Delta,\quad q_{\vec{c}}=\prod_{i=1}^n q_{c_i}.
\end{split}
\eq
One can see from (\ref{eq:coprod}) that the $n$-coproduct of $E(z)$ consists of $n$ factors as follows:
\bq
\begin{split}
\Delta^{(n-1)}(E(z))=&E(C^{-1}_2C^{-1}_3\cdots C_n^{-1}z)\otimes K^+(C_2^{-1}C_3^{-1}\cdots C_n^{-1}z)\otimes\cdots\otimes K^+(C_n^{-1}z)\\
&+1\otimes E(C^{-1}_3\cdots C_n^{-1}z)\otimes K^+(C_3^{-1}\cdots C_n^{-1}z)\otimes\cdots\otimes K^+(C_n^{-1}z)\\
&+\cdots\\
&+1\otimes\cdots\otimes E(z).
\end{split}
\label{eq:Ezcopro}
\eq
The representation $\rho_{\vec c}$ on the tensor product Fock space $\mathcal{F}_{\vec{c}}(\vec{u})$ is written as the sum of $n$ vertex operators:
\begin{align}
\rho_{\vec c}\left(E(z)\right)&=\sum_{i=1}^n\frac{1-q_{c_i}}{\kappa_1}\Lambda_i(z),\label{eq:Ecop}\\
\Lambda_i(z)&=\underbrace{1\otimes\cdots\otimes1}_{i-1}\otimes \eta_{c_i}(q_{c_{i+1}}^{-1/2}\cdots q_{c_n}^{-1/2}z)\otimes\varphi_{c_{i+1}}^+(q_{c_{i+1}}^{-1/2}\cdots q_{c_n}^{-1/2}z)\otimes\cdots\otimes\varphi_{c_n}^+(q_{c_n}^{-1/2}z).
\end{align}
For later convenience, we adjust the normalization parameter $u_i$ to rewrite it as,
\ba
&&\rho_{\vec c}\left(E(z)\right)=\sum_{i=1}^ny_i\Lambda_i(z),\qquad y_i=\frac{q_{c_i}^{1/2}-q_{c_i}^{-1/2}}{q_3^{1/2}-q_3^{-1/2}}.\label{eq:DIMnormalization}
\ea
We note that the $\Lambda_i(z)$ obtained here are identical to the bosonic representation which we proposed in (\ref{Lambda_i}) if we identify $u_i=\exp\left(\epsilon\frac{h_1h_2h_3}{h_{c_i}^2}a_0^{(i)}\right)$.
For reference, we write down the contraction between $\Lambda_i(z)$ and $\Lambda_j(w)$ again:
\ba
\Lambda_i(z)\Lambda_j(w)=\begin{dcases}
\exp\left(\sum_{r=1}^\infty\frac{\kappa_r}{r(1-q_{c_i}^r)}\biggl(\frac{w}{z}\biggr)^r\right):\Lambda_i(z)\Lambda_j(w):\quad(i=j)\\
\ \ :\Lambda_i(z)\Lambda_j(w):\quad(i>j)\\
\exp\left(\sum_{r=1}^\infty\frac{\kappa_r}{r}\biggl(\frac{w}{z}\biggr)^r\right):\Lambda_i(z)\Lambda_j(w):\quad(i<j)\ \ .
\end{dcases}
\label{eq:tLcont}
\ea

When we need to eliminate the extra $\mathfrak{gl}_1$ factor as in the quantum $W_N$, we should use $t(z)$ instead of $E(z)$. 
We can obtain the coproduct of $\alpha(z),\beta(z)$  just by replacing $H_{\pm r}$ and $C$ with $\Delta(H_{\pm r})$ and $\Delta(C)$. Because it consists of a single vertex operator, $t(z)$ also acts as the sum of $n$ vertex operators,
\ba
\rho_{\vec c}\left(t(z)\right)=\sum_{i=1}^ny_i\tilde{\Lambda}_i(z),
\ea
where $\tilde{\Lambda}(z)$ matches again the vertex operator we obtained in (\ref{eq:Lambda_tilde}).

\paragraph{An alternative identification with Drinfeld currents: }
One can also use the Drinfeld current $F(z)$ to describe the Miura operators. Instead of (\ref{eq:Ecop}), we write
\begin{align}
	&\rho_{\vec{c}}(F(z))=\sum_{i=1}^{n}y_{i}\Lambda_{i}^{*}(z),\label{eq:Fcop}
\end{align}
where
\begin{align}
	&\Lambda_{i}^{*}(z)=\varphi^{-}_{c_{1}}(q_{c_{1}}^{-1/2}z)\otimes\cdots\otimes\varphi^{-}_{c_{i-1}}(q_{c_{1}}^{-1/2}\cdots q_{c_{i-1}}^{-1/2}z)\otimes\xi_{i}(q_{c_{i}}^{-1/2}\cdots q_{c_{i-1}}^{-1/2}z)\underbrace{\otimes1\cdots\otimes 1}_{n-i}.
\end{align}
The contraction formula (\ref{eq:tLcont}) are modified to, 
\begin{align}
	&\Lambda_{i}^{*}(z)\Lambda_{j}^{*}(w)=\begin{dcases}\exp\left(\sum_{r=1}^{\infty}-\frac{\kappa_{r}}{r(1-q_{c_{i}}^{-r})}\right):\Lambda_{i}^{*}(z)\Lambda_{j}^{*}(w):\quad(i=j),\\:\Lambda_{i}^{*}(z)\Lambda_{j}^{*}(w):\quad(i>j),\\\exp\left(\sum_{r=1}^{\infty}-\frac{\kappa_{r}}{r}\left(\frac{w}{z}\right)^{r}\right):\Lambda_{i}^{*}(z)\Lambda_{j}^{*}(w):\quad(i<j)\end{dcases} \,.
\end{align}
For this identification, the definition of the Miura operators should be modified\footnotemark
\begin{align}
	R^{*(c)}(z)=\sum_{n=0}^{\infty}:\prod_{k=1}^{n}\left(-\frac{q_{c}^{1/2}q_{3}^{-(k-1)/2}-q_{c}^{-1/2}q_{3}^{(k-1)/2}}{q_{3}^{k/2}-q_{3}^{-k/2}}\Lambda^{*}(q_{3}^{k-1}z)\right):q_{3}^{nD_{z}}\quad(c=1,2,3).
\end{align}
\footnotetext{One can indeed obtain the Miura operators and contraction formula for $F(z)$ by using the Miki automorphism (\ref{eq:Mikiduality}) directly on the formula derived from $E(z)$. After using Miki automorphism the central element changes as $C\rightarrow C^{-1}$ so we need to reverse all the parameters as $q_{c}\rightarrow q_{c}^{-1}$.}

\subsection{Comparison with \texorpdfstring{$q$}{q}-deformed Miura transformation}
In this section, we establish the direct relation between the quantum toroidal $\mathfrak{gl}_1$ and the $q$-deformed corner VOA defined by the Miura transformation (\ref{qMiura}). We first use the first identification of the Drinfeld currents with the Miura operator (\ref{eq:Ecop}),

\begin{screen}
We claim that the product of the Drinfeld currents:
\ba
E_m(z)\equiv E(q_{3}^{-m+1}z)E(q_{3}^{-m+2}z)\cdots\cdots E(z),
\label{eq:EEEE}
\ea
in $\rho_{\vec c}$ representation can be identified with the current generated by the Miura transformation $T_m(z)$:
\ba
R_1^{(c_{1})}R_2^{(c_{2})}\cdots R_n^{(c_{n})}=\sum_{m=0}^{\infty}(-1)^{m}T_{m}(z)q_{3}^{-mD_{z}}.
\ea
when it is operated on the Fock space $\mathcal{F}_{c_1}\otimes\mathcal{F}_{c_2}\otimes\cdots\otimes\mathcal{F}_{c_n}$. Namely,
\begin{align}
	\rho_{\vec{c}}(E_{m}(z))=T_{m}(z).
\end{align}
It establishes that the Miura transformation we proposed properly describes the $q$-deformed corner VOA.\footnotemark
\end{screen}
\footnotetext{One may wonder why the shift parameter in ({\ref{eq:EEEE}}) is set to $q_3$ although there is triality symmetry among $q_1,q_2,q_3$. This is because we defined the Miura operator so that $R^{(3)}$ would consist of finite terms.}
In the rest of this section, we give a proof of this statement.
From  (\ref{eq:tLcont}), one can see that $\Lambda_{i}(q_{3}^{-1}z)\Lambda_{j}(z)$ vanishes when $i<j$. Then we have
\small
\ba
E_m(z)=\sum_{\substack{m_1,\cdots,m_n\geq0\\ \sum_{i=1}^n m_i=m}}\  \overrightarrow{\prod_{i_n=1}^{m_n}}y_n\Lambda_n(q_3^{-m+i_n}z) \overrightarrow{\prod_{i_{n-1}=1}^{m_{n-1}}}y_{n-1}\Lambda_{n-1}(q_3^{-m+m_n+i_{n-1}}z)\cdots\overrightarrow{\prod_{i_1=1}^{m_1}}y_1\Lambda_1(q_3^{-m_1+i_1}z).
\ea
\normalsize
The product symbol with an arrow implies a fixed ordering of factors:
\ba
\overrightarrow{\prod_{i_j=1}^{m_j}}\Lambda_j(q_3^{-(m_1+\cdots+m_j)+i_j}z)\equiv\Lambda_j(q_3^{-(m_1+\cdots+m_j)+1}z)\Lambda_j(q_3^{-(m_1+\cdots+m_j)+2}z)\cdots\Lambda_j(q_3^{-(m_1+\cdots+m_{j-1})}z).
\ea
To confirm the above claim, we only have to check that we can reproduce the coefficient  in (\ref{qMiuraOp})  by rewriting (\ref{eq:EEEE}) with the normal ordered product.
The contraction between the vertex operators can be evaluated from the following formula derived from (\ref{eq:tLcont}):
\begin{eqnarray}
\Lambda_{j}(q_{3}^{-\ell}z)\Lambda_{k}(z)&=&
\begin{cases} 
\chi_j^{(\ell)}:\Lambda_{j}(q_{3}^{-\ell}z)\Lambda_{k}(z): & (j=k) \\
:\Lambda_{j}(q_{3}^{-\ell}z)\Lambda_{k}(z): & (j>k),
\end{cases}
\end{eqnarray}
where we set
\begin{eqnarray}
&&\chi_{j}^{(\ell)}=\frac{(1-q_{3}^{\ell})(1-q_{3}^{\ell}q_{c_{j}}^{-1})}{(1-q_{c_{j}+1}q_{3}^{\ell})(1-q_{c_{j}-1}q_{3}^{\ell})}.
\end{eqnarray}
Using this relation, we have 
\ba
\overrightarrow{\prod_{i_j=1}^{m_j}}y_j\Lambda_j(q_3^{-(m_1+\cdots+m_j)+i_j}z)=y_j^{m_j}\prod_{\ell=1}^{m_j-1}(\chi_j^{(\ell)})^{m_j-\ell}:\prod_{i_j=1}^{m_j}\Lambda_j(q_3^{-(m_1+\cdots+m_j)+i_j}z):.
\ea
When  $c_j\neq3$, we have $\chi_j^{(\ell)}=\frac{(1-q_{3}^{\ell})(1-q_{3}^{\ell}q_{c_{j}}^{-1})}{(1-q_{3}^{\ell+1})(1-q_{3}^{\ell-1}q_{c_{j}}^{-1})}$  and
\ba
y_j^{m_j}\prod_{\ell=1}^{m_j-1}(\chi_j^{(\ell)})^{m_j-\ell}=\prod_{\ell=1}^{m_j}\frac{q_{c_j}^{1/2}q_{3}^{(1-\ell)/2}-q_{c_j}^{-1/2}q_{3}^{-(1-\ell)/2}}{q_{3}^{\ell/2}-q_{3}^{-\ell/2}}.
\label{eq:DIM_Miura}
\ea
One can check that this expression is also true for $c_j=3$ because both hand sides indeed become zero for $m_j\geq2$.
The factor (\ref{eq:DIM_Miura})  is exactly the same as the coefficient in (\ref{qMiuraOp}). Thus, $E_m(z)$ matches $T_m(z)$.\qed

One may obtain the similar statement for the second identification with the Drinfeld currents (\ref{eq:Fcop}).
Defining the product of Drinfeld currents as
\begin{align}
	F_{m}(z)&\equiv F(q_{3}^{m-1}z)F(q_{3}^{m-2}z)\cdots F(z)
\end{align}
and currents generated from Miura transformation as
\ba
R_1^{*(c_{1})}R_2^{*(c_{2})}\cdots R_n^{*(c_{n})}\equiv\sum_{m=0}^{\infty}(-1)^{m}T_{m}^{*}(z)q_{3}^{mD_{z}},
\ea
the correspondence between the two is
\begin{align}
	\rho_{\vec{c}}(F_{m}(z))=T_{m}^{*}(z).
\end{align}
The proof of these claims is exactly parallel to the first identification.

Before finishing this section, we comment on the order independence of the Miura operators. As we mentioned, one expects that the corner VOA $Y_{L, M, N}$ does not depend on the order of the Miura operators in the Miura transformation.  The proof of the statement was difficult since handling the higher currents was difficult. 

In the $q$-deformed version, we can provide a straightforward proof of this claim.  In this section, we have established a direct link between the Miura transformation in the order $\vec c$ with the $\rho_{\vec c}$ representation of the quantum toroidal $\mathfrak{gl}_1$. We can map the ordering independence of the Miura transformation to that of the tensor product representation $\rho_{\vec c}$.  In the quantum toroidal $\mathfrak{gl}_1$, there is a universal R-matrix\footnote{The explicit form the R-matrix in terms of the free boson is rather complicated even for the tensor product of the form $\mathcal{F}_a\otimes\mathcal{F}_a$ \cite{garbali2020r} (see also \cite{Fukuda:2017qki,Awata:2016mxc}).}. One can show the equivalence between the tensor products of two Fock representations with a different order, $\mathcal{F}_{c}\otimes\mathcal{F}_{c'}$ and $\mathcal{F}_{c'}\otimes\mathcal{F}_c$, by using this R-matrix. This fact, after coming back to the equivalent Miura transformation, implies the ordering independence of Miura operators.

\subsection{Screening currents }
In this section, we discuss the screening charges of the $q$-deformed $Y$ algebra defined by the Miura transformation.
The screening charges for the quantum toroidal $\mathfrak{gl}_1$ realized on $\mathcal{F}_{c_1}\otimes\mathcal{F}_{c_2}\otimes\cdots\otimes\mathcal{F}_{c_n}$ were constructed in  \cite{bershtein2018plane} as the vertex operators commuting with the Drinfeld currents. 
Because we have already expressed all the generators determined by the Miura transformation as the product of the Drinfeld currents, the screening charges in \cite{bershtein2018plane} work for the $q$-deformed $Y$ algebra as well. To make this paper self-contained, we give their explicit form below. 

Let us first consider the product of the two Fock spaces $\mathcal{F}_c(u_1)\otimes\mathcal{F}_c(u_2)$ of the same type.  There are two screening charges as follows:
\begin{align}
S^\pm_{cc}&=\oint dzS^\pm_{cc}(z),\\
S^+_{cc}(z)&=e^{\frac{h_{c+1}}{h_c}Q^-}z^{-\frac{h_{c+1}}{h_c}a^-_0+\frac{h_c}{h_{c-1}}}\exp\left(\sum_{r=1}^\infty\frac{-(q_{c+1}^{r/2}{-}q_{c+1}^{-r/2})}{r(q_c^{r/2}{-}q_c^{-r/2})}v_{-r}z^r\right)\exp\left(\sum_{r=1}^\infty\frac{(q_{c+1}^{r/2}{-}q_{c+1}^{-r/2})}{r(q_c^{r/2}{-}q_{c}^{-r/2})} v_{r}z^{-r}\right),\\
S^-_{cc}(z)&=e^{\frac{h_{c-1}}{h_c}Q^-}z^{-\frac{h_{c-1}}{h_c}a^-_0+\frac{h_c}{h_{c+1}}}\exp\left(\sum_{r=1}^\infty\frac{-(q_{c-1}^{r/2}{-}q_{c-1}^{-r/2})}{r(q_c^{r/2}{-}q_c^{-r/2})}v_{-r}z^r\right)\exp\left(\sum_{r=1}^\infty\frac{(q_{c-1}^{r/2}{-}q_{c-1}^{-r/2})}{r(q_c^{r/2}{-}q_{c}^{-r/2})} v_{r}z^{-r}\right),
\end{align}
where 
\ba
&&v_r=q_c^{r/2} a_r^{(1)}-q_c^{r} a_r^{(2)},\quad v_{-r}=q_c^{-r} a_{-r}^{(1)}-q_c^{-r/2} a_{-r}^{(2)} \qquad (r>0),\\
&&a_0^-=a_0^{(1)}-a_0^{(2)},\quad Q^-=Q^{(1)}-Q^{(2)}.
\ea
The operator $Q$ is defined as $[a_n,Q]=\frac{-h_c^3}{\sigma}\delta_{n,0}$.
This is exactly the screening charges for $q$-Virasoro algebra \cite{Shiraishi:1995rp}. For different types of the Fock spaces, say, $\mathcal{F}_1(u_1)\otimes\mathcal{F}_2(u_2)$, the screening charge is given by
\begin{equation}
	\begin{split}
		S_{12}&=\oint S_{12}(z)\mathrm{d}z,\\
		S_{12}(z)&=e^{\frac{h_2}{h_1} Q^{(1)}-\frac{h_1}{h_2} Q^{(2)}}z^{\frac{h_2}{h_1} a_0^{(1)}-\frac{h_1}{h_2} a_0^{(2)}+\frac{h_2}{h_3}}\exp\left(\sum_{r=1}^\infty \frac{1}{-r} v_{-r}'z^r\right)\exp\left(\sum_{r=1}^\infty \frac{1}{r} v_{r}'z^{-r}\right),
	\end{split}
\end{equation}
where 
\bq
\begin{split}
	v_{-r}'&=\frac{q_1^{-r}(q_2^{r/2}-q_2^{-r/2})}{q_1^{r/2}-q_1^{-r/2}}a_{-r}^{(1)}-\frac{q_1^{-r/2}(q_1^{r/2}-q_1^{-r/2})}{q_2^{r/2}-q_2^{-r/2}} a_{-r}^{(2)}, \\
	v_r'&=\frac{q_1^{r/2}(q_2^{r/2}-q_2^{-r/2})}{q_1^{r/2}-q_1^{-r/2}}a_r^{(1)}-\frac{q_3^{-r/2}(q_1^{r/2}-q_1^{-r/2})}{q_2^{r/2}-q_2^{-r/2}} a_r^{(2)}, 
	\qquad (r>0). 
\end{split}
\eq
For a generic  tensor product $\mathcal{F}_{c_1}(u_1)\otimes\mathcal{F}_{c_2}(u_2)\otimes\cdots\otimes\mathcal{F}_{c_n}(u_n)$, the screening charge $S_{c_ic_{i+1}}$ is assigned to each pair  $\mathcal{F}_{c_i}(u_i)\otimes\mathcal{F}_{c_{i+1}}(u_{i+1})$ of neighboring Fock spaces.
The commutativity with the Drinfeld currents can be immediately seen from the coproduct structure such  as  (\ref{eq:Ezcopro}).

Since the screening charges commute with the Drinfeld currents, they also commute with the generators derived from Miura operators, which are equivalent to the product of the Drinfeld currents. We conclude that the screening currents for the Miura operators are the same as the algebra defined in \cite{bershtein2018plane}. 

For reference, we write down the explicit commutation relation between the screening currents and the higher generators. There are three nontrivial cases: $\mathcal{F}_1(u_1)\otimes\mathcal{F}_3(u_2)$, $\mathcal{F}_1(u_1)\otimes\mathcal{F}_2(u_2)$ and $\mathcal{F}_1(u_1)\otimes\mathcal{F}_1(u_2)$.
For the first case, we have
\begin{align}
    \left[T_{l}(z), S_{13}(w)\right]&=\prod_{j=1}^{l-1}\frac{(1-j;1)}{(j;0)}q_{3}^{\frac{l}{2}}\left(\delta(\frac{w}{q_{3}^{-l}z})-q_{2}\delta(\frac{w}{q_{3}^{-l}q_{2}^{-1}z})\right):\prod_{j=1}^{l}\Lambda_{1}(q_{3}^{-j+1}z)S_{13}(q_{3}^{-l}z):\nonumber\\
    &=\prod_{j=1}^{l-1}\frac{(1-j;1)}{(j;0)}q_{3}^{\frac{l}{2}}\mathfrak{D}_{q_{2}}\left(w\delta(\frac{w}{q_{3}^{-l}z}):\prod_{j=1}^{l}\Lambda_{1}(q_{3}^{-j+1}z)S_{13}(q_{3}^{-l}z):\right),
\end{align}
where we used the notation
\begin{equation}
    \mathfrak{D}_{a}f(w)=\frac{f(w)-f(wa)}{w},\qquad
(p;q)=q_{3}^{\frac{p}{2}}q_{1}^{\frac{q}{2}}-q_{3}^{-\frac{p}{2}}q_{1}^{-\frac{q}{2}}.
\label{eq:totalderivative}
\end{equation}
We note that $(p;q)$ is not the $q$-Pochhammer symbol.
For the second case, we have 
\begin{align}
&[T_{l}(z),S_{12}(w)]\nonumber\\
&=\sum_{n=0}^{l}\sum_{q=1}^{l-n}q_{1}^{-\frac{l}{2}}q_{3}^{-\frac{n}{2}}\notag\\
&\quad\times\prod_{j=1}^{n+q}\left(-\frac{(j-1;-1)}{(j;0)}\right)\prod_{k=1}^{l-n-q}\left(-\frac{(k;1)}{(k;0)}\right)\left(\prod_{j\neq n+1}^{n+q}\frac{(j-n;1)}{(j-n-1;0)}\right)(1;1)\left(\prod_{k=1}^{l-n}\frac{(k+q;1)}{(k+q;0)} \right)\nonumber\\
&\quad\times q_{3}^{-q}\mathfrak{D}_{q_{3}^{q}}\left(w\delta(\frac{w}{q_{1}q_{3}^{-n+1}z}):\prod_{j=1}^{n}\Lambda_{1}(q_{3}^{-j+1}z)\prod_{k=1}^{l-n}\Lambda_{2}(q_{3}^{-k+1-n}z)S_{12}(q_{1}q_{3}^{-q-n+1}z):\right).
\end{align}
For the last case, we have two screening currents $S_{11}^\pm(w)$. The commutation relations are as follows:
\begin{align}
&\left[T_{l}(z),S_{11}^{+}(w)\right]=-\sum_{n=0}^{l}\sum_{q=1}^{l-n}(q_{1}^{\frac{1}{2}}q_{3}^{\frac{1}{2}})^{l-2n}\nonumber\\
&\quad\times\prod_{j=1}^{n}\frac{(1-j;1)}{(j;0)}\prod_{k=1}^{l-n}\frac{(1-k;1)}{(k;0)}\left(\prod_{j=1}^{n}\frac{(j-n-q;1)}{(j-n-q-1;0)}\right)\left(\prod_{k\neq q}^{l-n}\frac{(k-q-1;-1)}{(k-q;0)}\right)(1;1)\nonumber\\
&\quad\times q_{3}^{-q}\mathfrak{D}_{q_{3}^{q}}\left(w\delta(\frac{w}{q_{3}^{\frac{1}{2}-n}z}):\prod_{j=1}^{n}\Lambda_{1}(q_{3}^{-j+1}z)\prod_{k=1}^{l-n}\Lambda_{2}(q_{3}^{-n-k+1}z)S_{11}^{+}(q_{3}^{-n-q+\frac{1}{2}}z):\right),\\
&\left[T_{l}(z),S^{-}_{11}(w)\right]
=\sum_{n=1}^{l}q_{3}^{n-\frac{l}{2}}(0;1)\prod_{p=1}^{n-1}\frac{(-p;1)}{(p;0)}\prod_{p=1}^{l-n}\frac{(-p;1)}{(p;0)}\nonumber\\
&\quad\times\mathfrak{D}_{q_{2}}\left(w\delta\left(\frac{w}{q_{1}^{-\frac{1}{2}}q_{3}^{-n+\frac{1}{2}}z}\right)
:\prod_{j=1}^{n-1}\Lambda_{1}(q_{3}^{-j+1}z)\prod_{k=1}^{l-n+1}\Lambda_{2}(q_{3}^{-n-k+2}z)S_{11}^{-}(q_{1}^{\frac{1}{2}}q_{3}^{-n+\frac{3}{2}}z):\right).
\end{align}
As is expected, all the above relations are expressed in the form of total difference.

\section{Summary and future directions}
In this paper, we proposed the Miura transformation associated with the $q$-deformed corner vertex operator algebra. It is based on the $q$-deformed version of the fractional power differential operators, which can be expressed by $q$-Pochhammer symbol and the vertex operators. We show that the operator thus defined has much simpler properties compared with the undeformed case.
\begin{itemize}
	\item The higher generators have a manageable form than the undeformed case. Indeed, it can be identified with the products of the Drinfeld currents in the direct product representations of the quantum toroidal $\mathfrak{gl}_1$.
	\item The commutativity of the screening operators with the higher currents is difficult to prove analytically for the undeformed case. After $q$-deformation, one can identify them with those of the toroidal algebra. The proof of the commutativity with the higher currents becomes a straightforward consequence of the known results.
	\item  We can prove the order independence of the Miura operators from the existence of the universal R-matrix of the toroidal algebra.
\end{itemize}
In a sense, the symmetry property (Miki automorphism) of the toroidal algebra helps give a simple understanding of the whole picture. The appearance of the fractional power in the undeformed case finds a natural origin in the rewriting of the Drinfeld currents in a Fock space $\mathcal{F}_c$ with the coefficients with a different parameter.

There are a few subjects that we wish clarify in the near future.
\begin{itemize}
	\item Proof of the quadratic relation (\ref{eq:fTT_pred}): we have conjectured a quadratic relation for the $q$-deformed corner vertex operator algebra. We give a partial proof of the statement in appendix \ref{ap:quadratic}. It will be desirable to complete the proof.\footnote{While preparing this draft, we noticed a paper \cite{Kojima2021}, which seems to be very relevant to this question.}. We are also interested in the relation between the quadratic basis of $W_{1+\infty}$ in \cite{Prochazka:2014gqa} and the primary basis of $q$-deformed corner VOA.
	\item Matrix analog of the Miura transformation: in \cite{Eberhardt:2019xmf,Rapcak:2019wzw}, the authors proposed a matrix generalization of the Miura transformation to describe the rectangle $W$ algebra (see for instence, \cite{Creutzig:2018pts, Creutzig:2019qos}). We conjecture that its $q$-deformation can be directly related to the quantum toroidal $\mathfrak{gl}_k$ (see for instance, \cite{feigin2013representations,Negut:2020}).
	\item The computation of the $R$ matrix for the mixed Miura operators:  while the existence of the universal R-matrix is known, the free boson representation of such R-matrix is very relevant for the explicit computation. While it was conjectured in \cite{Prochazka2019}, it does not capture the deformation parameters appearing in the higher currents and may need some modifications. In the $q$-deformed case, the R-matrix is given in \cite{garbali2020r} for the bosonic case. We hope to derive the corresponding formula for the fermionic cases.
	\item The relation between the $q$-deformation and M-theory:
	the authors of \cite{Gaiotto2019,Gaiotto2020} found that the Y-algebra appeared also in the system of M2-branes and M5-branes.
	It may be interesting to explore how the $q$-deformation of corner VOA in our paper can be interpreted in terms of M-theory.
\end{itemize}

\section*{Acknowledgement}
We would like to thank Tomas Prochazka for discussions.
We are grateful to the anonymous referee for detailed and useful comments.
KH is supported in part by JSPS fellowship. YM is supported in part by JSPS KAKENHI Grant numbers 18K03610 and 18H01141. GN is supported in part by FoPM, the University of Tokyo. AW is supported in part by JSPS fellowship, MEXT and JSR Fellowship, the University of Tokyo.

\appendix
\section{Quadratic Relations}\label{ap:quadratic}
In this section, we give a partial proof of the quadratic relations, which we conjectured in (\ref{eq:fTT_pred}). The analysis is limited to the case where one of the generators is the lowest one $T_{1}(z)$. The complete proof would be obtained by using the fusion formulas in \cite{Kojima2019}, which we will not pursue here. We claim,
\begin{screen}
\begin{align}\label{eq:FTTlowest}
    &f_{1,m}\left(q_{3}^{-\frac{m-1}{2}}\frac{w}{z}\right)T_{1}(z)T_{m}(w)-f_{m,1}\left(q_{3}^{\frac{m-1}{2}}\frac{z}{w}\right)T_{m}(w)T_{1}(z)\nonumber\\
    =&\frac{(q_{1}^{\frac{1}{2}}-q_{1}^{-\frac{1}{2}})(q_{2}^{\frac{1}{2}}-q_{2}^{-\frac{1}{2}})}{(q_{3}^{\frac{1}{2}}-q_{3}^{-\frac{1}{2}})}\left[\delta\left(q_{3}\frac{w}{z}\right)T_{m+1}(q_{3}w)-\delta\left(q_{3}^{-m}\frac{w}{z}\right)T_{m+1}(w) \right]
\end{align}
\end{screen}
In this section we consider the general Fock representation $\mathcal{F}_{c_1}(u_1)\otimes\mathcal{F}_{c_2}(u_2)\otimes\cdots\otimes\mathcal{F}_{c_n}(u_n)$. We  also assume that the number of Fock spaces of type 1, 2, 3 is not be the same (namely $q_{\vec{c}}\neq 1$), such that one can eliminate the $\mathfrak{gl}_{1}$ factor. 
For later convenience, we rewrite (\ref{eq:tildeLambda_no}) and (\ref{eq:def_fij}) as follows,   
\begin{align}
    f_{1,1}(z)=\exp\left(-\sum_{r=1}^{\infty}\frac{\kappa_{r}}{r}\frac{(q_{3}^{r}-q_{\vec{c}}^{r})}{(1-q_{3}^{r})(1-q_{\vec{c}}^{r})}z^{r}\right),
\end{align}
\begin{align}\label{eq:FTTtildeLambda_no}
    f_{1,1}(\frac{w}{z})\tilde{\Lambda}_i(z)\tilde{\Lambda}_j(w)=\begin{dcases}
\gamma_{c_{i}}(\frac{w}{z}):\tilde{\Lambda}_i(z)\tilde{\Lambda}_j(w):\quad(i=j),\\
\Delta(q_{3}^{-\frac{1}{2}}\frac{w}{z}):\tilde{\Lambda}_i(z)\tilde{\Lambda}_j(w):\quad(i>j),\\
\Delta(q_{3}^{\frac{1}{2}}\frac{w}{z}):\tilde{\Lambda}_i(z)\tilde{\Lambda}_j(w):\quad(i<j),
\end{dcases}
\end{align}
where we set
\begin{align}
    &\Delta(z)=\frac{(1-q_{1}q_{3}^{\frac{1}{2}}z)(1-q_{1}^{-1}q_{3}^{-\frac{1}{2}}z)}{(1-q_{3}^{\frac{1}{2}}z)(1-q_{3}^{-\frac{1}{2}}z)},\\
    &\gamma_{c_{i}}(z)=\frac{(1-q_{c_{i}}z)(1-q_{c_{i}}^{-1}z)}{(1-q_{3}z)(1-q_{3}^{-1}z)}.
\end{align}
We note  $\gamma_{3}(z)=1$.
We obtain from (\ref{eq:def_fij})
\begin{align}
    f_{1,i}(z)=\frac{\prod_{k=1}^{i}f_{1,1}(q_{3}^{\frac{1}{2}(i+1-2k)}z)}{\prod_{k=1}^{i-1}\Delta(q_{3}^{\frac{1}{2}(i-2k)}z)}.
\end{align}
For simplicity, we use the notation
\begin{align}
    T_{m}(z)=\sum_{\sum_{i=1}^{n}m_{i}=m}\prod_{k=1}^{n}A(m_{k},c_{k}):\prod_{l=1}^{n}(\prod_{j_{l}=1}^{m_{l}}\tilde{\Lambda}_{l}(q_{3}^{-(\sum_{k=1}^{l-1}m_{k}+j_{l}-1)}z)):,
\end{align}
where we set
\begin{align*}A(m_{k},c_{k})\equiv\prod_{j_{k}=1}^{m_{k}}\frac{q_{c_{k}}^{\frac{1}{2}}q_{3}^{\frac{1}{2}(1-j_{k})}-q_{c_{k}}^{-\frac{1}{2}}q_{3}^{-\frac{1}{2}(1-j_{k})}}{q_{3}^{\frac{j_{k}}{2}}-q_{3}^{-\frac{j_{k}}{2}}}.\end{align*}
The left-hand side of (\ref{eq:FTTlowest}) is
\begin{align}
    &f_{1,m}\left(q_{3}^{-\frac{m-1}{2}}\frac{w}{z}\right)T_1(z)T_{m}(w)-f_{m,1}\left(q_{3}^{\frac{m-1}{2}}\frac{z}{w}\right)T_{m}(w)T_1(z)\nonumber\\
    =&\frac{(q_{1}^{\frac{1}{2}}-q_{1}^{-\frac{1}{2}})(q_{2}^{\frac{1}{2}}-q_{2}^{-\frac{1}{2}})}{(q_{3}^{\frac{1}{2}}-q_{3}^{-\frac{1}{2}})}\sum_{p=1}^{n}\sum_{m_{1}+\cdots+m_{n}=m}\prod_{k\neq p}A(m_{k},c_{k}) A(m_{p}+1,c_{p})\nonumber\\
    &\times\left(\delta\left(\frac{z}{q_{3}^{-(m_{1}+\cdots+m_{p-1})+1}w}\right)-\delta\left( \frac{z}{q_{3}^{-(m_{1}+\cdots+\cdots m_{p})}w}\right) \right):\tilde{\Lambda}_{p}(z)\prod_{l=1}^{n}(\prod_{j_{l}=1}^{m_{l}}\tilde{\Lambda}_{l}(q_{3}^{-(\sum_{i=1}^{l-1}m_{i}+j_{l}-1)}w)):.
\end{align}
The right-hand side contains many terms which are not in  (\ref{eq:FTTlowest}), but the terms not containing $\delta\left(\frac{z}{q_{3}w}\right)$ or $\delta\left(\frac{z}{q_{3}^{-m}w}\right)$ actually cancel out each other. 
To see that, let us divide it into four parts as follows:
\begin{itemize}
    \item $p=1$ of the first delta function 
    \begin{align}
        \sum_{\sum_{i=1}^{n}m_{i}=m}\prod_{k\neq1}A(m_{k},c_{k})A(m_{1}+1,c_{1}):\tilde{\Lambda}_{1}(q_{3}w)\prod_{l=1}^{n}(\prod_{j_{l}=1}^{m_{l}}\tilde{\Lambda}_{l}(q_{3}^{-(\sum_{k=1}^{l-1}m_{k}+j_{l}-1)}w)):\delta\left(\frac{z}{q_{3}w}\right)\label{eq:FTTlowest1}
    \end{align}
    \item $2\leq p \leq n$ of the first delta function
    \begin{align}
        &\sum_{p=2}^{n}\sum_{\sum_{i=1}^{n}m_{i}=m}\prod_{k\neq p}A(m_{k},c_{k})A(m_{p+1},c_{p})\nonumber\\
        &\quad\quad\times:\tilde{\Lambda}_{p}(q_{3}^{-\sum_{j=1}^{p-1}m_{j}+1}w)\prod_{l=1}^{n}(\prod_{j_{l}=1}^{m_{l}}\tilde{\Lambda}_{l}(q_{3}^{-(\sum_{k=1}^{l-1}m_{k}+j_{l}-1)}w)):\delta\left(\frac{z}{q_{3}^{-\sum_{j=1}^{p-1}m_{j}+1}w}\right)\label{eq:FTTlowest2}
    \end{align}
    \item $1\leq p\leq n-1$ of the second delta function
    \begin{align}
        &-\sum_{p=1}^{n-1}\sum_{\sum_{i=1}^{n}m_{i}=m}\prod_{k\neq p}A(m_{k},c_{k})A(m_{p}+1,c_{p})\nonumber\\
        &\quad\quad\times:\tilde{\Lambda}_{p}(q_{3}^{-\sum_{j=1}^{p}m_{j}}w)\prod_{l=1}^{n}(\prod_{j_{l}=1}^{m_{l}}\tilde{\Lambda}_{l}(q_{3}^{-(\sum_{k=1}^{l-1}m_{k}+j_{l}-1)}w)):\delta\left(\frac{z}{q_{3}^{-\sum_{j=1}^{p}m_{j}}w}\right)\label{eq:FTTlowest3}
    \end{align}
    \item $p=n$ of the second delta function
    \begin{align}
        &-\sum_{\sum_{i=1}^{n}m_{i}=m}\prod_{k\neq n}A(m_{k},c_{k})A(m_{n}+1,c_{n})\notag\\
        &\quad\quad \times:\prod_{l=1}^{n}(\prod_{j_{l}=1}^{m_{l}}\tilde{\Lambda}_{l}(q_{3}^{-(\sum_{k=1}^{l-1}m_{k}+j_{l}-1)}w))\tilde{\Lambda}_{n}(q_{3}^{-m}w):\delta\left(\frac{z}{q_{3}^{-m}w}\right)\label{eq:FTTlowest4}
    \end{align}
\end{itemize}
In the following, we prove that the terms proportional to $\delta\left(\frac{z}{q_{3}^{-r+1}w}\right) (1\leq r\leq m)$ vanish and the remaining terms are given by
\begin{align}
    &\text{(\ref{eq:FTTlowest2})}+\text{(\ref{eq:FTTlowest3})}\nonumber\\
    =&\sum_{p=2}^{n}\sum_{m_{p}+\cdots+m_{n}=m}\prod_{k\neq p}A(m_{k},c_{k})A(m_{p}+1,c_{p}):\tilde{\Lambda}_{p}(q_{3}w)\prod_{l=p}^{n}(\prod_{j_{l}=1}^{m_{l}}\tilde{\Lambda}_{l}(q_{3}^{-(\sum_{k=p}^{l-1}m_{k}+j_{l}-1)}w)):\delta\left(\frac{z}{q_{3}w} \right)\nonumber\\
    &-\sum_{p=1}^{n-1}\sum_{\sum_{i=1}^{p}m_{i}=m}\prod_{k\neq p}A(m_{k},c_{k})A(m_{p}+1,c_{p}):\prod_{l=1}^{p}(\prod_{j_{l}=1}^{m_{l}}\tilde{\Lambda}_{l}(q_{3}^{-(\sum_{k=1}^{l-1}m_{k}+j_{l}-1)}w))\tilde{\Lambda}_{p}(q_{3}^{-m}w):\delta\left(\frac{z}{q_{3}^{-m}w}\right). \label{eq:FTTlowest2+3}
\end{align}

The coefficients of $\delta\left(\frac{z}{q_{3}^{-r+1}w}\right) (1\leq r\leq m)$  are
\begin{align}
    \sum_{p=2}^{n}\left\{\sum_{\substack{\sum_{i=1}^{p-1}m_{i}=r\\\sum_{i=p}^{n}=m-r}}\prod_{k\neq p}A(m_{k},c_{k})A(m_{p}+1,c_{p}):\tilde{\Lambda}_{p}(q_{3}^{-r+1}w)\prod_{l=1}^{n}(\prod_{j_{l}=1}^{m_{l}}\tilde{\Lambda}_{l}(q_{3}^{-(\sum_{k=1}^{l-1}m_{k}+j_{l}-1)}w)):\label{eq:FTTintermediate1}\right.\\
    \left.-\sum_{\substack{\sum_{i=1}^{p-1}m_{i}'=r-1\\\sum_{i=p}^{n}m_{i}=m-r+1}}\prod_{k\neq p-1}A(m_{k}',c_{k})A(m_{p-1}'+1,c_{p-1}):\tilde{\Lambda}_{p-1}(q_{3}^{-r+1}w)\prod_{l=1}^{n}(\prod_{j_{l}=1}^{m_{l}'}\tilde{\Lambda}_{l}(q_{3}^{-(\sum_{k=1}^{l-1}m_{k}'+j_{l}-1)}w)):\right\}.\label{eq:FTTintermediate2}
\end{align}

To derive (\ref{eq:FTTlowest2+3}), we need to prove that the sum of (\ref{eq:FTTintermediate1}) and (\ref{eq:FTTintermediate2}) vanishes.
Because (\ref{eq:FTTintermediate2}) contains at least one $\tilde{\Lambda}_{p-1}$, 
it is convenient to consider separately the case of $m_{p-1}=0$ and the case of $m_{p-1}\geq 1$ in (\ref{eq:FTTintermediate1}).
Similarly, we consider $m'_p=0$ and $m'_p\geq 1$ in (\ref{eq:FTTintermediate2}) separately.
Then we have
\begin{align}
    \text{(\ref{eq:FTTintermediate1})}=&\sum_{p=2}^{n}\left\{\sum_{\substack{m_{1}+\cdots+m_{p-2}=r\\m_{p}+\cdots+m_{n}=m-r\\m_{p-1}=0}}\prod_{k\neq p}A(m_{k},c_{k})A(m_{p}+1,c_{p})\right.\nonumber\\
    &\times:\tilde{\Lambda}_{p}(q_{3}^{-r+1}w)\prod_{l=1}^{n}(\prod_{j_{l}=1}^{m_{l}}\tilde{\Lambda}_{l}(q_{3}^{-(\sum_{k=1}^{l-1}m_{k}+j_{l}-1)}w)):\label{eq:FTTlast1}\\
    +&\sum_{\substack{m_{1}+\cdots+m_{p-2}=r\\m_{p}+\cdots+m_{n}=m-r\\m_{p-1}\geq1}}\prod_{k\neq p,p-1}A(m_{k},c_{k})A(m_{p}+1,c_{p})A(m_{p-1},c_{p-1})\nonumber\\
    &\times\left.:\tilde{\Lambda}_{p}(q_{3}^{-r+1}w)\prod_{l=1}^{n}(\prod_{j_{l}=1}^{m_{l}}\tilde{\Lambda}_{l}(q_{3}^{-(\sum_{k=1}^{l-1}m_{k}+j_{l}-1)}w)):\right\},\label{eq:FTTlast2}
\end{align}

\begin{align}
    \text{(\ref{eq:FTTintermediate2})}&=-\sum_{p=2}^{n}\left\{\sum_{\substack{\sum_{i=1}^{p-1}m_{i}'=r-1\\
    	\sum_{i=p+1}^{n}m_{i}=m-r+1\\m_{p}'=0}}\prod_{k\neq p-1}A(m_{k}',c_{k})A(m_{p-1}'+1,c_{p-1})\nonumber\right.\\
    &\quad\quad\times:\tilde{\Lambda}_{p-1}(q_{3}^{-r+1}w)\prod_{l=1}^{n}(\prod_{j_{l}=1}^{m_{l}'}\tilde{\Lambda}_{l}(q_{3}^{-(\sum_{k=1}^{l-1}m_{k}'+j_{l}-1)}w)):\label{eq:FTTlast3}\\
    &+\sum_{\substack{\sum_{i=1}^{p-1}m_{i}'=r-1\\\sum_{i=p}^{n}m_{i}=m-r+1\\m_{p}'\geq1}}\prod_{k\neq p-1,p}A(m_{k}',c_{k})A(m_{p-1}'+1,c_{p-1})A(m_{p}',c_{p})\nonumber\\
    &\left.\quad\quad\times:\tilde{\Lambda}_{p-1}(q_{3}^{-r+1}w)\prod_{l=1}^{n}(\prod_{j_{l}=1}^{m_{l}'}\tilde{\Lambda}_{l}(q_{3}^{-(\sum_{k=1}^{l-1}m_{k}'+j_{l}-1)}w)):\right\}.\label{eq:FTTlast4}
\end{align}
We can easily see (\ref{eq:FTTlast2})+(\ref{eq:FTTlast4})=0 under the identification
\begin{equation}
\begin{split}
    &m_{p-1}\geq1,\quad m_{p}'\geq1,\\
    &m_{i}=m_{i}' \quad(i\leq p-2),\\
    &m_{p-1}=m_{p-1}'+1,\\
    &m_{p}+1=m_{p}',\\
    &m_{i}=m_{i}'\quad (i\geq p+1).
\end{split}
\end{equation}

We can derive (\ref{eq:FTTlast1})+(\ref{eq:FTTlast3})=0  similarly.
The sum in (\ref{eq:FTTlast1}) is
\begin{align}\sum_{p=2}^{n}\sum_{\substack{m_{1}+\cdots+m_{p-2}=r\\m_{p}+\cdots+m_{n}=m-r\\m_{p-1}=0}}=\sum_{p=3}^{n}\sum_{\substack{m_{1}+\cdots+m_{p-2}=r\\m_{p}+\cdots+m_{n}=m-r\\m_{p-1}=0}},\end{align} because when $p=2$ there are no terms satisfying the condition $m_{1}+\cdots+m_{p-2}\geq1$. The same thing happens with (\ref{eq:FTTlast3}) and we get \begin{align}
\sum_{p=2}^{n}\sum_{\substack{\sum_{i=1}^{p-1}m_{i}'=r-1\\\sum_{i=p+1}^{n}m_{i}=m-r+1\\m_{p}'=0}}=\sum_{p=2}^{n-1}\sum_{\substack{\sum_{i=1}^{p-1}m_{i}'=r-1\\\sum_{i=p+1}^{n}m_{i}=m-r+1\\m_{p}'=0}}\\
=\sum_{p=3}^{n}\sum_{\substack{\sum_{i=1}^{p-2}m_{i}'=r-1\\\sum_{i=p}^{n}m_{i}=m-r+1\\m_{p-1}'=0}}
.\end{align} 
Applying the same argument, the region of $p$ decreases one by one and finally becomes zero. Then we obtain (\ref{eq:FTTlast1})+(\ref{eq:FTTlast3})=0 and (\ref{eq:FTTintermediate1})+(\ref{eq:FTTintermediate2})=0.\\

Summing (\ref{eq:FTTlowest1}), (\ref{eq:FTTlowest4}), (\ref{eq:FTTlowest2+3}),  we finally arrive at
\begin{align}
    &f_{1,m}\left(q_{3}^{-\frac{m-1}{2}}\frac{w}{z}\right)T_{1}(z)T_{m}(w)-f_{m,1}\left(q_{3}^{\frac{m-1}{2}}\frac{z}{w}\right)T_{m}(w)T_{1}(z)\nonumber\\
        =&\frac{(q_{1}^{\frac{1}{2}}-q_{1}^{-\frac{1}{2}})(q_{2}^{\frac{1}{2}}-q_{2}^{-\frac{1}{2}})}{(q_{3}^{\frac{1}{2}}-q_{3}^{-\frac{1}{2}})}\nonumber\\
    &\times\left(\sum_{p=1}^{n}\sum_{\substack{m_{p}+\cdots+m_{n}=m\\m_{1}=\cdots=m_{p-1}=0}}\prod_{k\neq p}A(m_{k},c_{k})A(m_{p}+1,c_{p}):\tilde{\Lambda}_{p}(q_{3}w)\prod_{l=p}^{n}(\prod_{j_{l}=1}^{m_{l}}\tilde{\Lambda}_{l}(q_{3}^{-(\sum_{k=1}^{l-1}m_{k}+j_{l}-1)}w)):\delta\left(\frac{z}{q_{3}w}\right)\nonumber\right.\\
    &\left.-\sum_{p=1}^{n}\sum_{\substack{m_{1}+\cdots m_{p}=m\\m_{p+1}=\cdots m_{n}=0}}\prod_{k\neq p}A(m_{k},c_{k})A(m_{p}+1,c_{p}):\prod_{l=1}^{p}(\prod_{j_{l}=1}^{m_{l}}\tilde{\Lambda}_{l}(q_{3}^{-(\sum_{k=1}^{l-1}m_{k}+j_{l}-1)}w))\tilde{\Lambda}_{p}(q_{3}^{-m}w):\delta\left(\frac{z}{q_{3}^{-m}w}\right)\right)\nonumber\\
    =&\frac{(q_{1}^{\frac{1}{2}}-q_{1}^{-\frac{1}{2}})(q_{2}^{\frac{1}{2}}-q_{2}^{-\frac{1}{2}})}{(q_{3}^{\frac{1}{2}}-q_{3}^{-\frac{1}{2}})}\left(T_{m+1}(q_{3}w)\delta\left(\frac{z}{q_{3}w}\right)-T_{m+1}(w)\delta\left(\frac{z}{q_{3}^{-m}w}\right)\right).
\end{align}
The last equation comes from the observation that the normal ordered vertex operator part of $T_{m+1}(z)$ can be written by adding one vertex operator to each side of the normal ordered vertex operator part of $T_{m}(z)$.

\bibliography{qW}

\providecommand{\href}[2]{#2}\begingroup\raggedright\begin{thebibliography}{10}

\bibitem{Alday2010}
L.~F. Alday, D.~Gaiotto, and Y.~Tachikawa, ``Liouville Correlation Functions
  from Four-dimensional Gauge Theories,''
  \href{http://dx.doi.org/10.1007/s11005-010-0369-5}{{\em Lett.Math.Phys.}
  {\bfseries 91} (2010) 167--197},
\href{http://arxiv.org/abs/0906.3219}{{\ttfamily arXiv:0906.3219 [hep-th]}}.

\bibitem{Tsymbaliuk:2014}
A.~Tsymbaliuk, ``The affine Yangian of gl$_1$ revisited,'' {\em Advances in
  Mathematics} {\bfseries 304} (2017) 583--645,
  \href{http://arxiv.org/abs/1404.5240}{{\ttfamily arXiv:1404.5240 [math.RT]}}.

\bibitem{Prochazka:2015deb}
T.~Proch\'{a}zka, ``{$ \mathcal{W} $ -symmetry, topological vertex and affine
  Yangian},'' \href{http://dx.doi.org/10.1007/JHEP10(2016)077}{{\em JHEP}
  {\bfseries 10} (2016) 077},
\href{http://arxiv.org/abs/1512.07178}{{\ttfamily arXiv:1512.07178 [hep-th]}}.

\bibitem{Gaiotto:2017euk}
D.~Gaiotto and M.~Rap\v{c}\'{a}k, ``{Vertex Algebras at the Corner},''
  \href{http://dx.doi.org/10.1007/JHEP01(2019)160}{{\em JHEP} {\bfseries 01}
  (2019) 160},
\href{http://arxiv.org/abs/1703.00982}{{\ttfamily arXiv:1703.00982 [hep-th]}}.

\bibitem{Prochazka:2017qum}
T.~Proch\'{a}zka and M.~Rap\v{c}\'{a}k, ``{Webs of W-algebras},''
  \href{http://dx.doi.org/10.1007/JHEP11(2018)109}{{\em JHEP} {\bfseries 11}
  (2018) 109},
\href{http://arxiv.org/abs/1711.06888}{{\ttfamily arXiv:1711.06888 [hep-th]}}.

\bibitem{bershtein2018plane}
M.~Bershtein, B.~Feigin, and G.~Merzon, ``Plane partitions with a “pit”:
  generating functions and representation theory,'' {\em Selecta Mathematica}
  {\bfseries 24} no.~1, (2018) 21--62,
  \href{http://arxiv.org/abs/1512.08779}{{\ttfamily arXiv:1512.08779 [math]}}.

\bibitem{Gaberdiel:2012ku}
M.~R. Gaberdiel and R.~Gopakumar, ``{Triality in Minimal Model Holography},''
  \href{http://dx.doi.org/10.1007/JHEP07(2012)127}{{\em JHEP} {\bfseries 1207}
  (2012) 127},
\href{http://arxiv.org/abs/1205.2472}{{\ttfamily arXiv:1205.2472 [hep-th]}}.

\bibitem{creutzig2020trialities}
T.~Creutzig and A.~R. Linshaw, ``Trialities of $\mathcal{W}$-algebras,''
  \href{http://arxiv.org/abs/2005.10234}{{\ttfamily arXiv:2005.10234
  [math.RT]}}.

\bibitem{Litvinov:2016mgi}
A.~Litvinov and L.~Spodyneiko, ``{On W algebras commuting with a set of
  screenings},'' \href{http://dx.doi.org/10.1007/JHEP11(2016)138}{{\em JHEP}
  {\bfseries 11} (2016) 138},
\href{http://arxiv.org/abs/1609.06271}{{\ttfamily arXiv:1609.06271 [hep-th]}}.

\bibitem{Prochazka:2018tlo}
T.~Proch\'{a}zka and M.~Rap\v{c}\'{a}k, ``{$ \mathcal{W} $ -algebra modules,
  free fields, and Gukov-Witten defects},''
  \href{http://dx.doi.org/10.1007/JHEP05(2019)159}{{\em JHEP} {\bfseries 05}
  (2019) 159},
\href{http://arxiv.org/abs/1808.08837}{{\ttfamily arXiv:1808.08837 [hep-th]}}.

\bibitem{Fateev1988}
V.~Fateev and S.~L. Lukyanov, ``{The Models of Two-Dimensional Conformal
  Quantum Field Theory with Z(n) Symmetry},''
\href{http://dx.doi.org/10.1142/S0217751X88000205}{{\em Int.J.Mod.Phys.}
  {\bfseries A3} (1988) 507}.

\bibitem{Eberhardt:2019xmf}
L.~Eberhardt and T.~Proch\'azka, ``{The matrix-extended $W_{1+\infty}$
  algebra},'' \href{http://dx.doi.org/10.1007/JHEP12(2019)175}{{\em JHEP}
  {\bfseries 12} (2019) 175}, \href{http://arxiv.org/abs/1910.00041}{{\ttfamily
  arXiv:1910.00041 [hep-th]}}.

\bibitem{Rapcak:2019wzw}
M.~Rap\v{c}\'{a}k, ``{On extensions of $
  \mathfrak{gl}\widehat{\left(\left.m\right|n\right)} $ Kac-Moody algebras and
  Calabi-Yau singularities},''
  \href{http://dx.doi.org/10.1007/JHEP01(2020)042}{{\em JHEP} {\bfseries 01}
  (2020) 042},
\href{http://arxiv.org/abs/1910.00031}{{\ttfamily arXiv:1910.00031 [hep-th]}}.

\bibitem{Shiraishi:1995rp}
J.~Shiraishi, H.~Kubo, H.~Awata, and S.~Odake, ``{A Quantum deformation of the
  Virasoro algebra and the Macdonald symmetric functions},''
  \href{http://dx.doi.org/10.1007/BF00398297}{{\em Lett. Math. Phys.}
  {\bfseries 38} (1996) 33--51},
\href{http://arxiv.org/abs/q-alg/9507034}{{\ttfamily arXiv:q-alg/9507034
  [q-alg]}}.

\bibitem{Awata:1995zk}
H.~Awata, H.~Kubo, S.~Odake, and J.~Shiraishi, ``{Quantum W(N) algebras and
  Macdonald polynomials},'' \href{http://dx.doi.org/10.1007/BF02102595}{{\em
  Commun. Math. Phys.} {\bfseries 179} (1996) 401--416},
  \href{http://arxiv.org/abs/q-alg/9508011}{{\ttfamily arXiv:q-alg/9508011}}.

\bibitem{Feigin:1995sf}
B.~Feigin and E.~Frenkel, ``{Quantum W algebras and elliptic algebras},''
  \href{http://dx.doi.org/10.1007/BF02108819}{{\em Commun. Math. Phys.}
  {\bfseries 178} (1996) 653--678},
  \href{http://arxiv.org/abs/q-alg/9508009}{{\ttfamily arXiv:q-alg/9508009}}.

\bibitem{Ding:1996mq}
J.-t. Ding and K.~Iohara, ``{Generalization and deformation of Drinfeld quantum
  affine algebras},''
\href{http://dx.doi.org/10.1023/A:1007341410987}{{\em Lett. Math. Phys.}
  {\bfseries 41} (1997) 181--193}.

\bibitem{Miki2007}
K.~Miki, ``{A (q, $\gamma$) analog of the $W_{1+\infty}$ algebra},''
  \href{http://dx.doi.org/10.1063/1.2823979}{{\em Journal of Mathematical
  Physics} {\bfseries 48} no.~12, (2007) 3520}.
  \url{http://scitation.aip.org/content/aip/journal/jmp/48/12/10.1063/1.2823979}.

\bibitem{Feigin:2015raa}
B.~Feigin, M.~Jimbo, T.~Miwa, and E.~Mukhin, ``{Quantum toroidal
  $\mathfrak{g}{{\mathfrak{l}}_{1}}$ and Bethe ansatz},''
  \href{http://dx.doi.org/10.1088/1751-8113/48/24/244001}{{\em J. Phys.}
  {\bfseries A48} no.~24, (2015) 244001},
\href{http://arxiv.org/abs/1502.07194}{{\ttfamily arXiv:1502.07194 [math.QA]}}.

\bibitem{Kojima2019}
T.~Kojima, ``Quadratic relations of the deformed $W$-superalgebra ${\cal W}_{q
  t}(\mathfrak{sl}(2|1))$,'' \href{http://arxiv.org/abs/1912.03096}{{\ttfamily
  arXiv:1912.03096 [math.QA]}}.

\bibitem{feigin1991duality}
B.~Feigin and E.~Frenkel, ``Duality in W-algebras,'' {\em International
  Mathematics Research Notices} {\bfseries 1991} no.~6, (1991) 75--82.

\bibitem{Prochazka:2014gqa}
T.~Proch\'{a}zka, ``{Exploring $ {\mathcal{W}}_{\infty } $ in the quadratic
  basis},'' \href{http://dx.doi.org/10.1007/JHEP09(2015)116}{{\em JHEP}
  {\bfseries 09} (2015) 116},
\href{http://arxiv.org/abs/1411.7697}{{\ttfamily arXiv:1411.7697 [hep-th]}}.

\bibitem{Gaberdiel:2011wb}
M.~R. Gaberdiel and T.~Hartman, ``{Symmetries of Holographic Minimal Models},''
  \href{http://dx.doi.org/10.1007/JHEP05(2011)031}{{\em JHEP} {\bfseries 1105}
  (2011) 031},
\href{http://arxiv.org/abs/1101.2910}{{\ttfamily arXiv:1101.2910 [hep-th]}}.

\bibitem{linshaw2017universal}
A.~R. Linshaw, ``Universal two-parameter $W_\infty$-algebra and vertex algebras
  of type $W(2, 3,..., N)$,'' \href{http://arxiv.org/abs/1710.02275}{{\ttfamily
  arXiv:1710.02275 [math.RT]}}.

\bibitem{schiffmann2013cherednik}
O.~Schiffmann and E.~Vasserot, ``{Cherednik algebras, W-algebras and the
  equivariant cohomology of the moduli space of instantons on A 2},'' {\em
  Publications math{\'e}matiques de l'IH{\'E}S} {\bfseries 118} no.~1, (2013)
  213--342, \href{http://arxiv.org/abs/1202.2756}{{\ttfamily arXiv:1202.2756
  [math]}}.

\bibitem{feigin2012quantum}
B.~Feigin, M.~Jimbo, T.~Miwa, E.~Mukhin, {\em et~al.}, ``Quantum toroidal
  $gl_1$-algebra: Plane partitions,'' {\em Kyoto Journal of Mathematics}
  {\bfseries 52} no.~3, (2012) 621--659,
  \href{http://arxiv.org/abs/1110.5310}{{\ttfamily arXiv:1110.5310}}.

\bibitem{Prochazka2019}
T.~Proch{\'{a}}zka, ``{Instanton R-matrix and $\mathcal{W}$-symmetry},''
  \href{http://dx.doi.org/10.1007/jhep12(2019)099}{{\em Journal of High Energy
  Physics} {\bfseries 2019} no.~12, (Dec, 2019) }.

\bibitem{Maulik:2012wi}
D.~Maulik and A.~Okounkov, ``{Quantum Groups and Quantum Cohomology},''
\href{http://arxiv.org/abs/1211.1287}{{\ttfamily arXiv:1211.1287 [math.AG]}}.

\bibitem{Zhu2015}
R.-D. Zhu and Y.~Matsuo, ``Yangian associated with 2D $\mathcal{N}= 1$
  {SCFT},'' \href{http://dx.doi.org/10.1093/ptep/ptv116}{{\em Progress of
  Theoretical and Experimental Physics} {\bfseries 2015} no.~9, (Sep, 2015)
  093A01}.

\bibitem{Andrews1986}
G.~Andrews, {\em Q-series : their development and application in analysis,
  number theory, combinatorics, physics, and computer algebra}.
\newblock Published for the Conference Board of the Mathematical Sciences by
  the American Mathematical Society, Providence, RI, 1986.

\bibitem{Ding1999}
J.~Ding and B.~Feigin, ``Quantized W-algebra of ${\frak sl}(2,1)$ : a
  construction from the quantization of screening operators,'' {\em Contemp.
  Math. 248, 83-108} (1999) ,
  \href{http://arxiv.org/abs/math/9801084}{{\ttfamily arXiv:math/9801084
  [math.QA]}}.

\bibitem{FHSSY:2010}
B.~Feigin, A.~Hoshino, J.~Shibahara, J.~Shiraishi, and S.~Yanagida, ``{Kernel
  function and quantum algebras},''
\href{http://arxiv.org/abs/1002.2485}{{\ttfamily arXiv:1002.2485 [math]}}.

\bibitem{garbali2020r}
A.~Garbali and J.~de~Gier, ``The $ R $-matrix of the quantum toroidal algebra $
  U_{q, t}(gl_1)$ in the Fock module,''
  \href{http://arxiv.org/abs/2004.09241}{{\ttfamily arXiv:2004.09241
  [math.QA]}}.

\bibitem{Fukuda:2017qki}
M.~Fukuda, K.~Harada, Y.~Matsuo, and R.-D. Zhu, ``{The
  Maulik\textendash{}Okounkov R-matrix from the
  Ding\textendash{}Iohara\textendash{}Miki algebra},''
  \href{http://dx.doi.org/10.1093/ptep/ptx123}{{\em PTEP} {\bfseries 2017}
  no.~9, (2017) 093A01}, \href{http://arxiv.org/abs/1705.02941}{{\ttfamily
  arXiv:1705.02941 [hep-th]}}.

\bibitem{Awata:2016mxc}
H.~Awata, H.~Kanno, A.~Mironov, A.~Morozov, A.~Morozov, Y.~Ohkubo, and
  Y.~Zenkevich, ``{Toric Calabi-Yau threefolds as quantum integrable systems. $
  \mathrm{\mathcal{R}} $ -matrix and $
  \mathrm{\mathcal{R}}\mathcal{T}\mathcal{T} $ relations},''
  \href{http://dx.doi.org/10.1007/JHEP10(2016)047}{{\em JHEP} {\bfseries 10}
  (2016) 047}, \href{http://arxiv.org/abs/1608.05351}{{\ttfamily
  arXiv:1608.05351 [hep-th]}}.

\bibitem{Kojima2021}
T.~Kojima, ``Quadratic relations of the deformed $W$-superalgebra
  $W_{qt}(A(M,N))$,'' \href{http://arxiv.org/abs/2101.01110}{{\ttfamily
  arXiv:2101.01110 [math.QA]}}.

\bibitem{Creutzig:2018pts}
T.~Creutzig and Y.~Hikida, ``{Rectangular W-algebras, extended higher spin
  gravity and dual coset CFTs},''
  \href{http://dx.doi.org/10.1007/JHEP02(2019)147}{{\em JHEP} {\bfseries 02}
  (2019) 147}, \href{http://arxiv.org/abs/1812.07149}{{\ttfamily
  arXiv:1812.07149 [hep-th]}}.

\bibitem{Creutzig:2019qos}
T.~Creutzig and Y.~Hikida, ``{Rectangular W algebras and superalgebras and
  their representations},''
  \href{http://dx.doi.org/10.1103/PhysRevD.100.086008}{{\em Phys. Rev. D}
  {\bfseries 100} no.~8, (2019) 086008},
  \href{http://arxiv.org/abs/1906.05868}{{\ttfamily arXiv:1906.05868
  [hep-th]}}.

\bibitem{feigin2013representations}
B.~Feigin, M.~Jimbo, T.~Miwa, and E.~Mukhin, ``Representations of quantum
  toroidal gl(n),'' {\em Journal of Algebra} {\bfseries 380} (2013) 78--108,
  \href{http://arxiv.org/abs/1204.5378}{{\ttfamily 1204.5378}}.

\bibitem{Negut:2020}
A.~Negut, ``{Deformed W-algebras in type A for rectangular nilpotent},''
\href{http://arxiv.org/abs/2004.02737}{{\ttfamily arXiv:2004.02737 [math.RT]}}.

\bibitem{Gaiotto2019}
D.~Gaiotto and J.~Oh, ``{Aspects of $\Omega$-deformed M-theory},''
  \href{http://arxiv.org/abs/1907.06495}{{\ttfamily arXiv:1907.06495
  [hep-th]}}.

\bibitem{Gaiotto2020}
D.~Gaiotto and M.~Rapcak, ``Miura operators, degenerate fields and the M2-M5
  intersection,'' \href{http://arxiv.org/abs/2012.04118}{{\ttfamily
  arXiv:2012.04118 [hep-th]}}.

\end{thebibliography}\endgroup
\bibliographystyle{utphys}
\end{document}